\documentclass[11pt]{article}
\usepackage{pifont}
\usepackage{natbib}
\usepackage{graphicx}
\usepackage{amssymb}
\usepackage{epstopdf}
\usepackage{amsmath}
\usepackage{subfigure}
\usepackage{lscape}
\usepackage{graphicx}
\usepackage{sidecap}
\usepackage{lineno}
\usepackage{setspace}
\usepackage{url}
\usepackage[hmargin=1in,vmargin=1in]{geometry}
\usepackage{authblk}
\usepackage{fancyhdr}
\usepackage[font={small}]{caption}

\begin{document}
\title{Tsunami penetration in tidal rivers, with observations of the Chile 2015 tsunami in rivers in Japan}
\author{Elena Tolkova}
\date{}
\maketitle
\begin{center}
NorthWest Research Associates

Bellevue, WA 98009-3027, USA \\

Tel.: +1425-889-5991 \\
Fax: +1425-556-9099 \\

e.tolkova@gmail.com, elena@nwra.com
\end{center}

The final publication is available at Springer via {\url{http://dx.doi.org/10.1007/s00024-015-1229-0}}
\tableofcontents

\begin{abstract}
An extensive data set of water level measurements of the September 2015 Chilean tsunami in rivers in Japan and a new methodology for data processing are used to verify that tsunami dissipation in a river at each instant and locality depends on the tidally-modified wave-locked slope of the river surface.
As deduced from the observations, 
a relatively small tsunami or ocean noise traveling at mild wave-locked slopes can propagate virtually without losses to the upstream locations where observed tidal ranges are a fraction of that downstream; though at the higher slopes, tidal and riverine currents combined efficiently damp the shorter waves. 
The observed correlations between the tsunami admittance upriver and the traveled wave-locked slopes are explained analytically under the fully-nonlinear shallow-water approximation. It is found that the wave-locked slope in a purely incident wave relates to the bottom drag in the same manner as a steady surface slope does for a stationary flow. For a small-amplitude tsunami in the study rivers, the wave-locked slope in a co-propagating tidal wave defines the background current and thereby friction experienced by the tsunami. 
\end{abstract}

\section{Introduction}

Recent mega-tsunami events of 2010 and 2011 demonstrated tsunami's superior ability to penetrate up rivers -- be they big or small, located in the near field or in the far field from the tsunami origin \citep{fritz2011, morisurvey, yeh, liu2013}. 
At the same time, tsunami in rivers exhibited behaviors not observed on an open coast. One such behavior is that tsunami propagation in coastal rivers, which are also invaded by ocean tides, is strongly influenced by tidal phase. Modeling studies, which had taken into account simultaneous propagation of tide and tsunami, showed that the present tsunami modeling framework of the shallow-water equations applied in initially still basins may not be sufficient to correctly represent the tsunami's propagation in tidal rivers \citep{zhang, tolkova, hill}. 

Analytically, tidal effects on tsunamis are attributed to the nonlinearity of the shallow-water equations (SWE).
Kowalik et al. (2006) and Kowalik and Proshutinsky (2010) investigated individual non-linear terms in the SWE as causes of interaction between tide and tsunami, and demonstrated the effects of each term on tsunami propagation in numerical simulations of hypothetical tsunami events in a tidal inlet. Relative contributions of different sources of nonlinearity, however, depend on a specific tidal environment. 
Nonlinearity of the river environments in particular is predominantly attributed to the quadratic bottom friction \citep{leblond, godin1999}.
Studies on interaction between tides and storm surges in shallow areas also point to bottom friction as the major contributor to the interaction among the flow components \citep{bernier2007, zhang2010}. Zhang W. et al (2010) interpreted the observed tidal modulation of the storm surge in the Taiwan Strait as variations of the surface slope along the channel needed to balance the increased force of bottom stress, to allow currents due to tide and the storm to flow through the Strait. 

Probably the first field observation of tidal influence on a tsunami in a river environment was made by Wilson and Torum in records of the 1964 Alaskan tsunami in the Columbia river. They noticed that 
the tsunami signal in Beaver at 86 rkm (river-kilometer, distance along a river from its mouth) and Vancouver at 170 rkm can only be detected atop high tide: ''Beaver tide gage, in particular, shows that, with the exception of the tsunami waves riding the tide crest, the intermediate waves have lost their identity and hardly register at low tide, though later waves are found again on the succeeding high tide'' \cite[]{wilson2}.
Kayane et al. (2011) found the same effect -- much higher tsunami signal at high tide than during the rest of the tidal cycle -- in records of the 2010 Chilean tsunami in several rivers in Japan. 
Yeh et al. (2012) described another effect found in the records of the 2011 Tohoku tsunami in the lower Columbia river: upon entering the river, the tsunami experienced excessive attenuation during the receding tide. This tendency was confirmed by means of numerical simulations \cite[]{tolkova}. For some reasons, the tsunami damping on ebb was clearly seen during the transition to the lower low water, but not during the second ebb of the mixed tide in the Columbia river estuary.
Tolkova et al. (2015) brought all the observations together to show that tsunami modulation by tide, in a manner in which receding tide dissipates tsunami the most, and the high tide dissipates it the least, is inherent to rivers in general; and suggested that the observed correlation of the tsunami attenuation rate with tidal phase is in fact correlation with tidally-modified slope of the river surface measured along a space-time trajectory of the tsunami, termed Wave-Locked Slope (WLS).

This work continues \cite[]{tolkova2015} with more detailed analysis of how ambient conditions affect far-field tsunami penetration in rivers, using the observations of the September 16, 2015 tsunami in five rivers on the Honshu east coast. 
Two-week-long continuous observations of shorter-period long waves (tsunami and ocean noise) in rivers will be used to analyze the effects of tides and river flow on propagation of these waves in a quantitative manner, as well as develop a methodology for such analysis.  The article is organized as follows. The study area and the collected data are described in the next section 2. Section 3, Preliminary Analysis of the Observations, explains physical quantities -- an instant wave amplitude and the Wave-Locked Slope -- used to quantify tsunami and ocean noise evolution and the river conditions, and demonstrates the correlation between the two, in yet another set of records.  Section 4, Admittance Computations, develops a methodology to compute the tsunami amplification/attenuation factor (admittance) as a function of WLS. The admittance factors are obtained for every station pair in every river and examined for mutual consistency and physical soundness. 
In the last section, we attempt to explain the physics behind WLS and its connection to the anticipated frictional interaction among the flow components. It is found that WLS in a purely incident wave relates to the bottom drag in the same manner as a steady surface slope does for a stationary flow. For a small-amplitude tsunami in the study rivers, WLS in a co-propagating tidal wave defines the background current and thereby friction experienced by the tsunami.
We conclude with discussing an important difference between dissipation of tide and dissipation of the shorter waves propagating atop tide, in a river. 

\section{Study area}

Two large 21th century trans-Pacific tsunamis originating offshore central Chile occurred on Feb. 27, 2010 after a Mw 8.8 earthquake and on Sep. 16, 2015 after a Mw 8.3 event. Approximately a day later, both tsunamis reached Japan. An underwater South Honshu Ridge, and the Kuril Islands chain direct trans-Pacific tsunamis onto the north-east coast of Honshu, and into its rivers. The Ministry of Land, Infrastructure, Transport, and Tourism (MLIT) of Japan maintains a network of water level stations in the rivers. Typically, the stations follow with a several km interval, with the first (most downstream) station located at about 1 km up the river mouth. Given the instrumentation and the geographical location, these rivers present an outstanding natural laboratory to study tsunami penetration upriver, especially with regard to relatively small, far-field tsunamis, which nowadays are fully controlled for by protective measures and can hardly make any damage there. The only complication comes from a relatively coarse (10-min) sampling interval in the measurements, originally intended for monitoring floods rather than tsunamis. 

\begin{figure}[ht]
\centering
	\resizebox{\textwidth}{!} %
		{\includegraphics{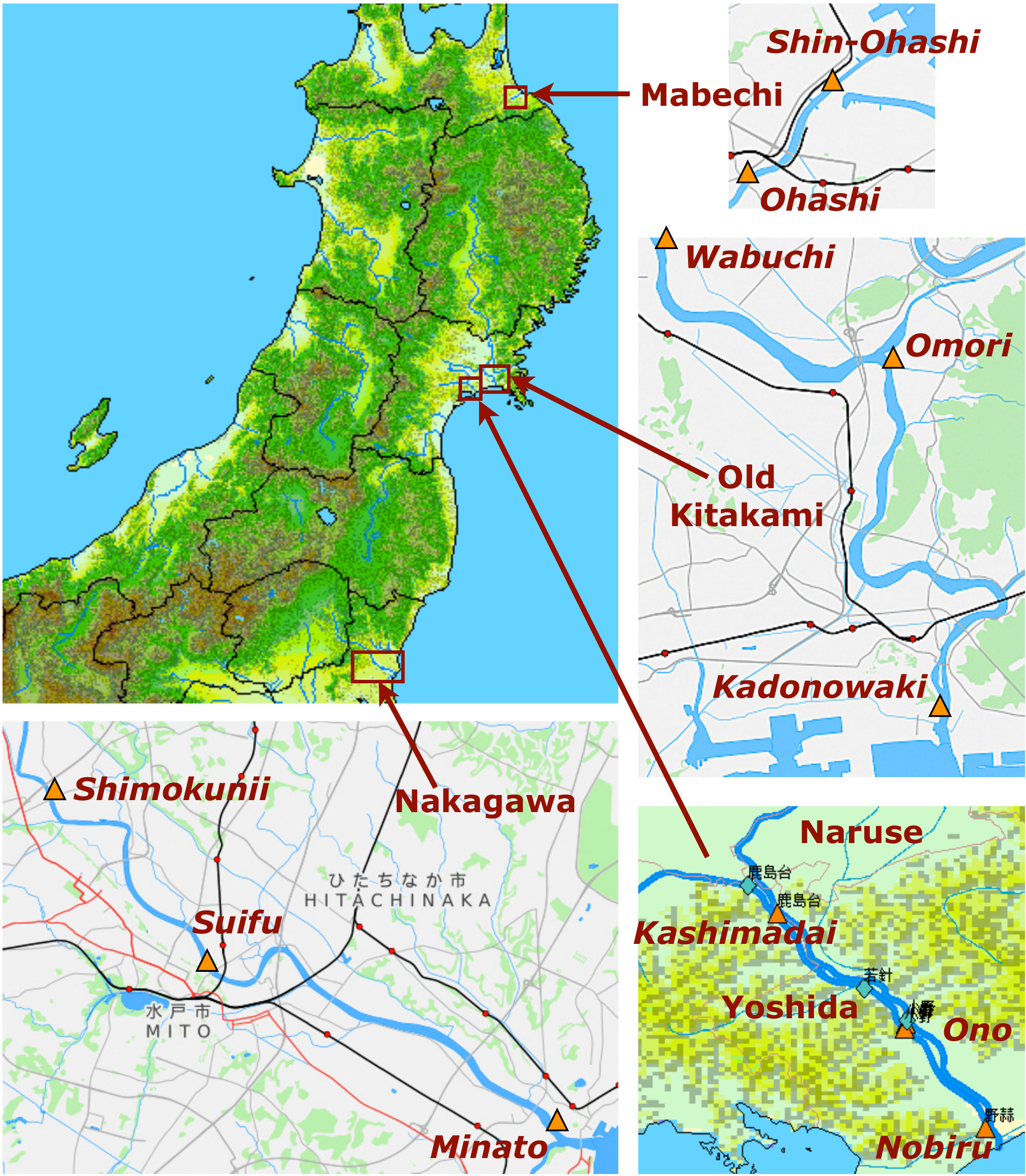}}
	\caption{North to South: Mabechi, Old Kitakami, Naruse/Yoshida, and Naka rivers on the Honshu east coast, and location of gauging stations in each river. Naruse is north of Yoshida. Map courtesy of MLIT and OpenStreetMap contributors. }
	\label{map}
\end{figure}

Both Chilean tsunamis penetrated several rivers and were recorded by MLIT gages. The records of the 2010 tsunami in Narise/Yoshida and Old Kitakami rivers can be found in \cite[]{tolkova2015}, while the records of the 2015 tsunami in these and two other rivers are presented below.
The 2015 tsunami, which arrived at the Honshu coast on Sep 18 at about 6:30 Japanese Standard Time (JST), was approximately 3 times lower than that in 2010, and 5 times higher than regular long-wave ocean noise. Nevertheless, the 2015 tsunami left traces at two gauging stations along Mabechi and Naruse, and at three stations along Old Kitakami, Yoshida, and Naka rivers (see Table 1 for the station list, and Figure \ref{map} for the map). 
\clearpage

\begin{figure}[ht]
\centering
	\resizebox{\textwidth}{!} %
		{\includegraphics{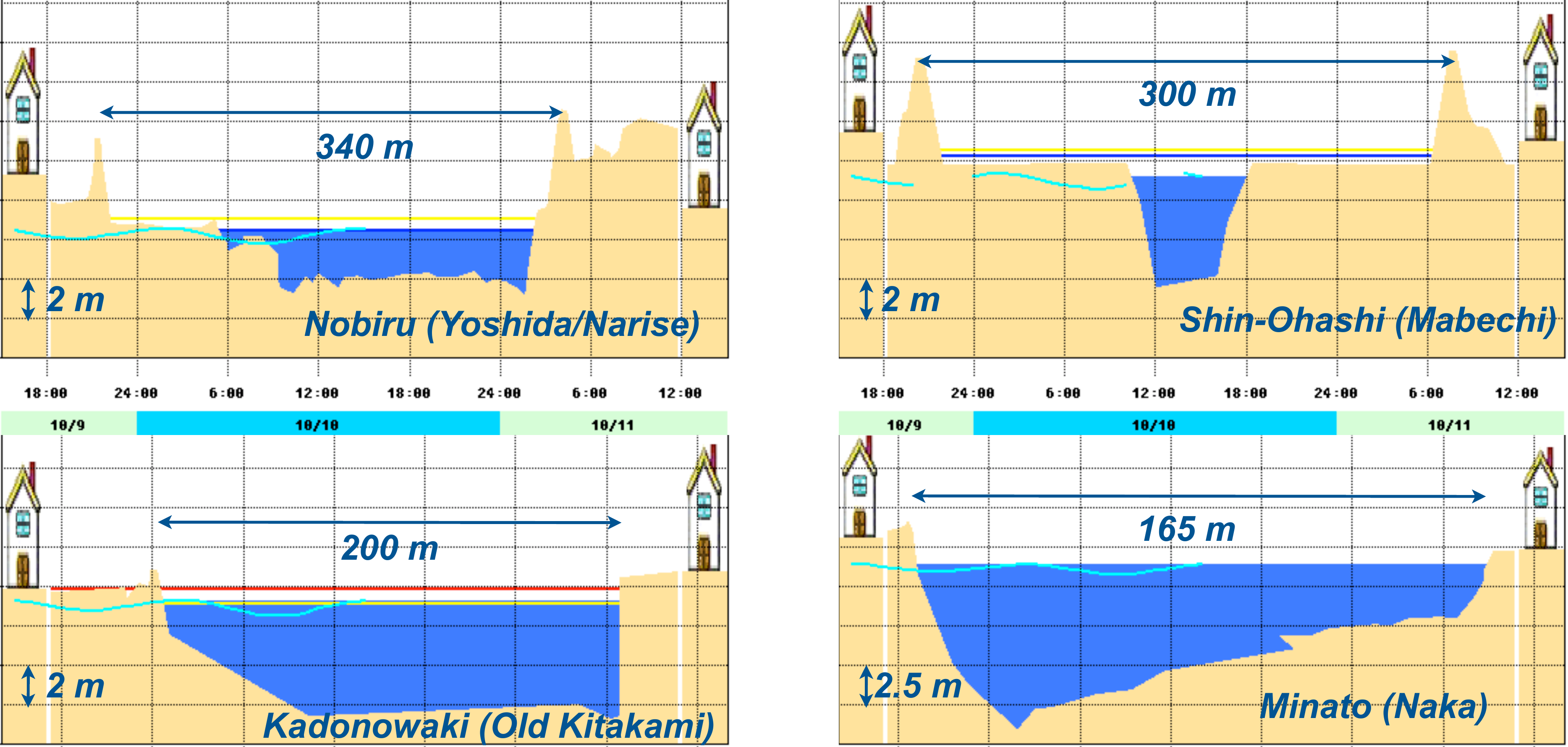}}  
	\caption{River profiles at their respective downstream gauging stations. Cyan curve - real-time water level on 10/10/2015. Image courtesy of MLIT.}
	\label{profiles}
\end{figure}

\begin{landscape}  
\begin{figure}[ht]  
\begin{tabular}{cc}
	\includegraphics[height=0.5\textwidth]{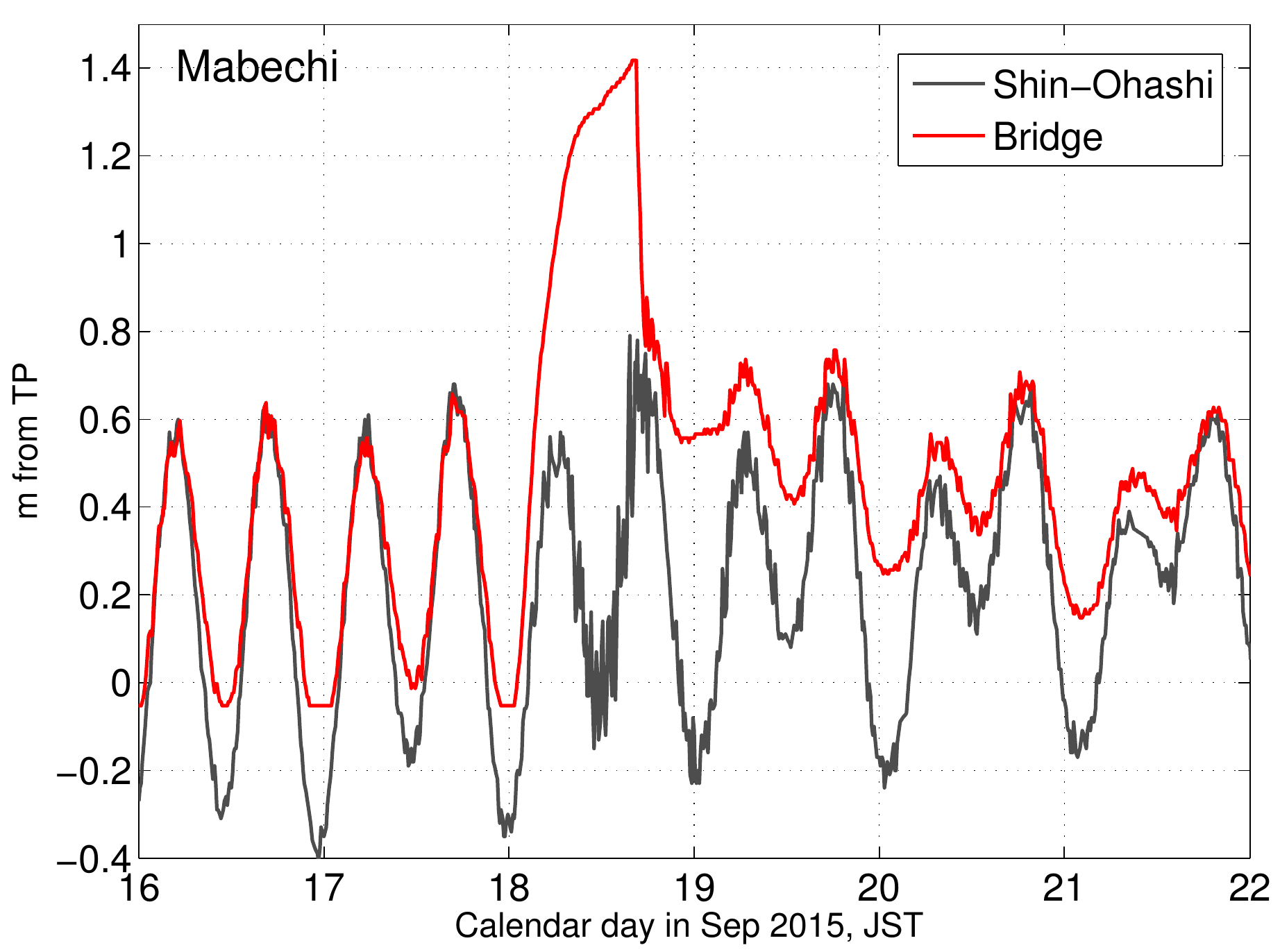} &
	\includegraphics[height=0.5\textwidth]{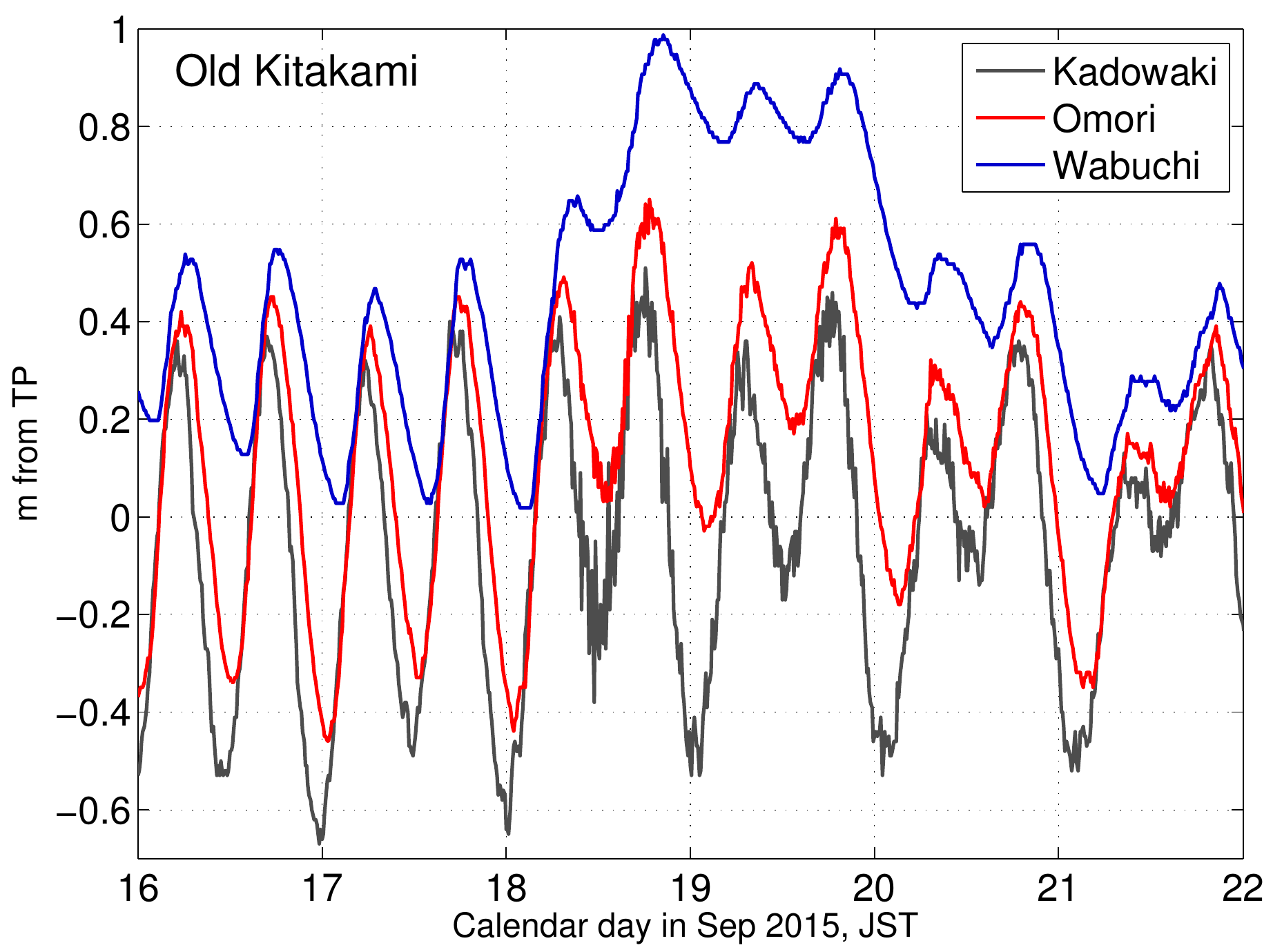} \\
	\includegraphics[height=0.5\textwidth]{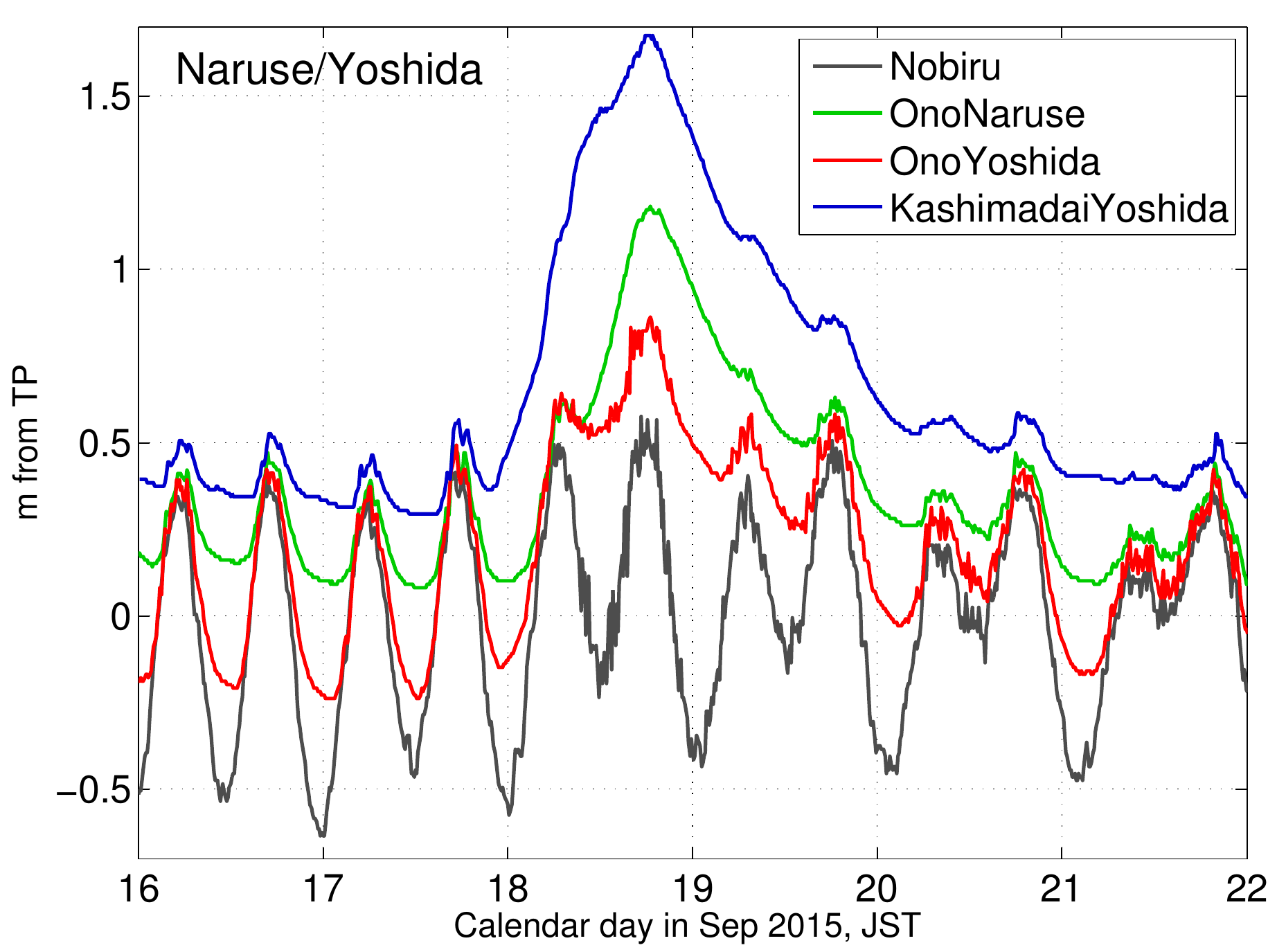} &
	\includegraphics[height=0.5\textwidth]{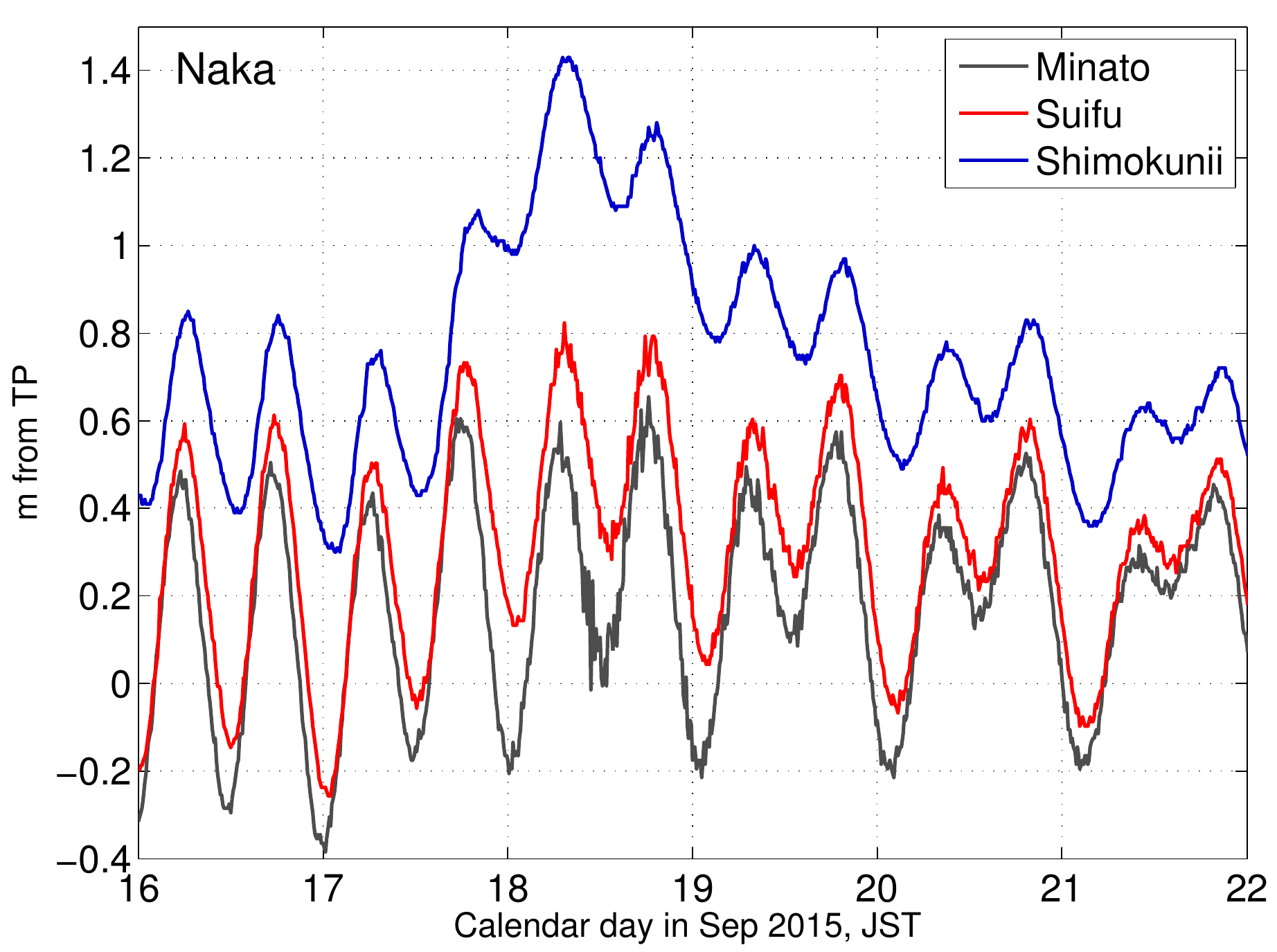}
\end{tabular}
	\caption{Water elevation records with traces of the 09/2015 Chilean tsunami event along Mabechi, Old Kitakami, Yoshida, and Naka rivers.}
	\label{recs}
\end{figure}
\end{landscape}  

For 9 km before the ocean, Naruse and Yoshida make two parallel channels separated by a narrow (50-200 m) strip of land; the rivers merge 0.9 km from the ocean. The first measurement station at Nobiru at 0.5 rkm is located in the common Naruse-Yoshida channel.  Upstream of the confluence, Narise is wider, shallower and steeper than Yoshida.
All five rivers have fortified banks continued with jetties at their mouthes, don't have wide estuaries or tidelands, and are much more channel-like than lowland continental rivers. The largest among these five rivers is the Naka river, followed by Old Kitakami. 
The river profiles at their respective downstream stations are shown in Figure \ref{profiles}.
 
The segments of the original 2+ week long records through the tsunami event in these rivers are shown in Figure \ref{recs}. The records contain the tsunami signal, as well as ever-present long-wave noise, likely originating with seiches in coastal waters penetrating into the rivers (see Supplementary Appendix I for discussing evidence of coastal normal oscillations in the records).
The water level measurements are sampled at a 10-min interval, whereas the tsunami spectrum is localized between 1.5 to 2.5 c/h (cycle/hour) or periods from 25 to 40 min, and the ocean noise generates oscillations at 40-120 min period (see Suppl. Appendix I). Hence the sampling rate is marginal for the tsunami representation, but still allows to convey useful information; and the records clearly show the tsunami signal. 
\begin{table}
\begin{tabular}{|c|c|c|c|} \hline
River & Station & $x$ (rkm) & $\tau$ (min) \\ \hline
Naruse &Nobiru & 0.5 & -\\
              &Ono& 4.18 & 10 \\ \hline
Mabechi &Shin-Ohashi  & 1.2 & -\\
               & Ohashi (Bridge) & 4.0 & 10 \\ \hline
                 &Nobiru & 0.5 & -\\
 Yoshida  & Ono & 4.04 & 10 \\
                & Kashimadai & 8.99 & 30 \\  \hline          
 Old   & Kadonowaki  & 1.24 & -\\ 
Kitakami               & Omori & 13.17 & 30 \\   
                & Wabuchi & 21.78 & 60 \\ \hline  
         & Minato  & 1.1 & -\\
Naka  & Suifu  & 12.39 & 20 \\
           & Shimokunii  & 19.71 & 50 \\  \hline    
\end{tabular}
\caption{List of rivers, water level stations, their distances from the river mouth, and wave travel times (in 10 min increments) from the most downstream station in each river.}         
\end{table}

\section{Preliminary Analysis of the Observations}

Tolkova et al (2015) suggested that a decrease rate of an instant tsunami amplitude along a river varies with the tidally-modified Wave-Locked Slope (WLS) of the river surface. The instant tsunami amplitude, and WLS were defined as follows. The water level measurements $\eta(t)+s(t)$ at $x$ rkm (river-kilometer, distance along a river from its mouth) can be separated into the background signal $\eta(t)$ comprised of the tide and mean flow, and the tsunami signal $s(t)$. 
Neglecting dispersion, the tsunami is thought of as a train of trackable wave elements corresponding to individual measurements in a time series $s(t)$.
Each wave element is characterized by its amplitude $\alpha(t)$, also referred to as an instant wave amplitude, given by an absolute value of a complex-valued signal
\begin{equation}
\alpha(t)=|s(t)+i \hat{s}(t)| ,
\label{U}
\end{equation}
where $\hat{s}(t)$ represents the Hilbert transform of a time series $s(t)$  (Tolkova et al, 2015). In particular, for $s(t)=a \cdot cos(\omega t)$, $\alpha(t)=a$. 

An average WLS  $\beta_{AB}(t)$ between two stations $A$ and $B$ along a river represents an average background surface gradient traveled by a tsunami wave element passing $A$ at time $t-\tau$ and arriving at $B$ at time $t$, $\tau$ being the tsunami travel time from $A$ to $B$. WLS can be estimated as 
\begin{equation}
\beta_{AB}(t)=\left( \eta_B(t)-\eta_A(t-\tau) \right)/(x_B-x_A)
\label{betaAB}
\end{equation}
where $x_A$ and $x_B$ are along-river distances of the locations; the surface elevations $\eta_{A,B}$ are measured from the same vertical datum.  
Without tide, WLS would coincide with the steady hydraulic gradient supporting the river flow. In the tidal river, WLS is continuously modified by tide. Tsunami wave elements propagating at different WLSs (corresponding to different tidal phases) experience different 'rivers', with different mean currents and different slopes of the free surface. Tolkova et al (2015) found correlation between the reduction of the wave element amplitude and the WLS in the element's path, in water level measurements of the 2010 Chilean tsunami in the Yoshida River, Japan. 
These findings agree well with the field evidence that the tsunami intrusion distance into a river correlates well to the riverbed slope \citep{tanaka2013, tanaka2014} -- a proxy for the surface slope when different rivers are being compared. 
The longer records of the 2015 Chilean tsunami at more riverine locations permit to examine this correlation with a higher level of accuracy and reliability. 
\begin{figure}[ht]
	\resizebox{\textwidth}{!} %
		{\includegraphics{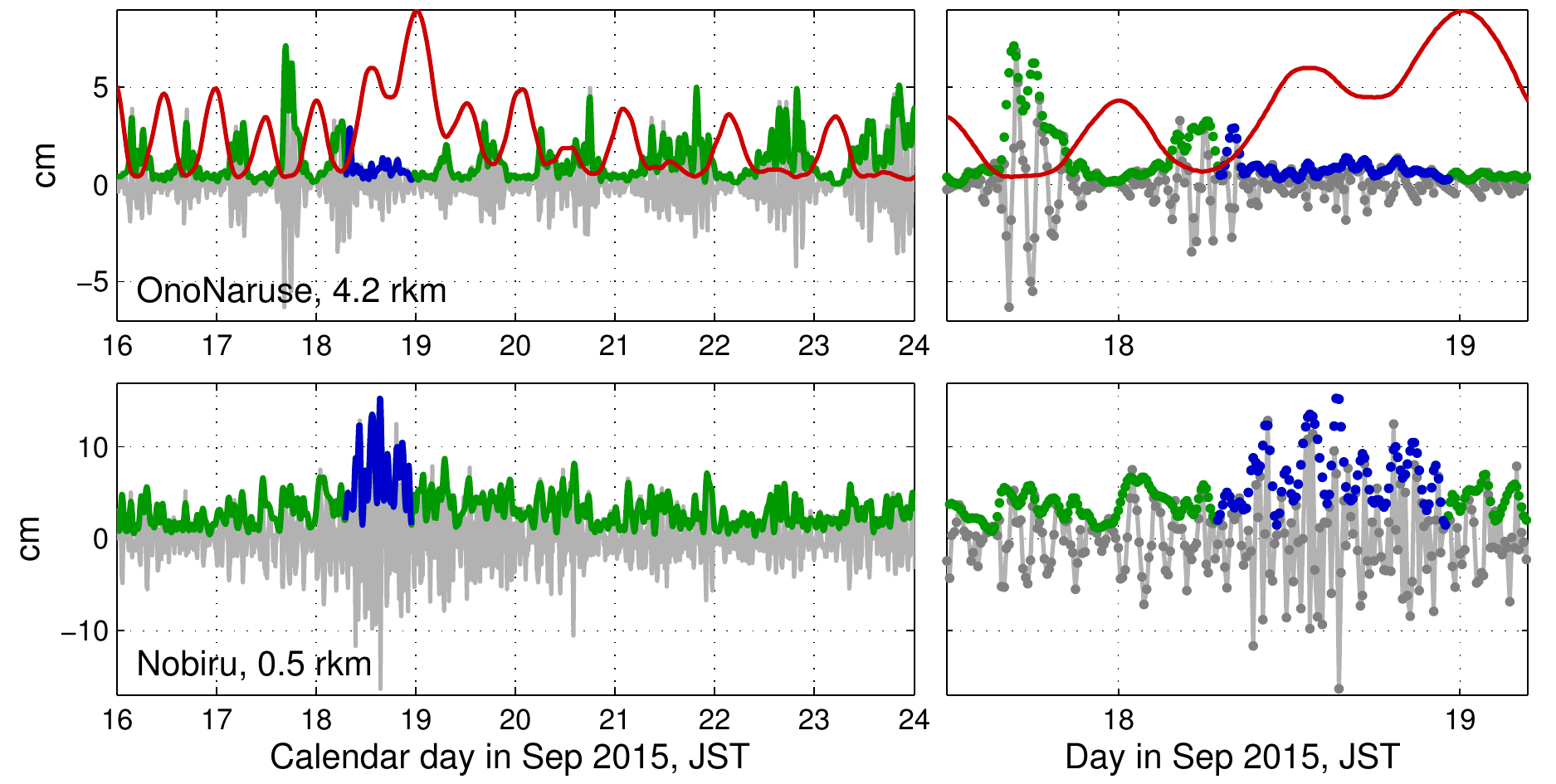}}  
	\caption{Segments of de-tided records at two stations in Naruse River, zoomed-in on the right: de-tided measurements (gray), amplitudes of the associated wave elements (green - due to long-wave ocean noise, and blue - for 16 h after the tsunami arrival at the coast), and wave-locked slope (red curve; in cm/rkm, with a factor 0.25) between the upriver station and the station near the river mouth (bottom panel).  Individual readings or wave elements are shown with dots in zoomed-in records on the right.}
	\label{wtimeNz}
\end{figure}
\begin{figure}[ht]
	\resizebox{\textwidth}{!} %
		{\includegraphics{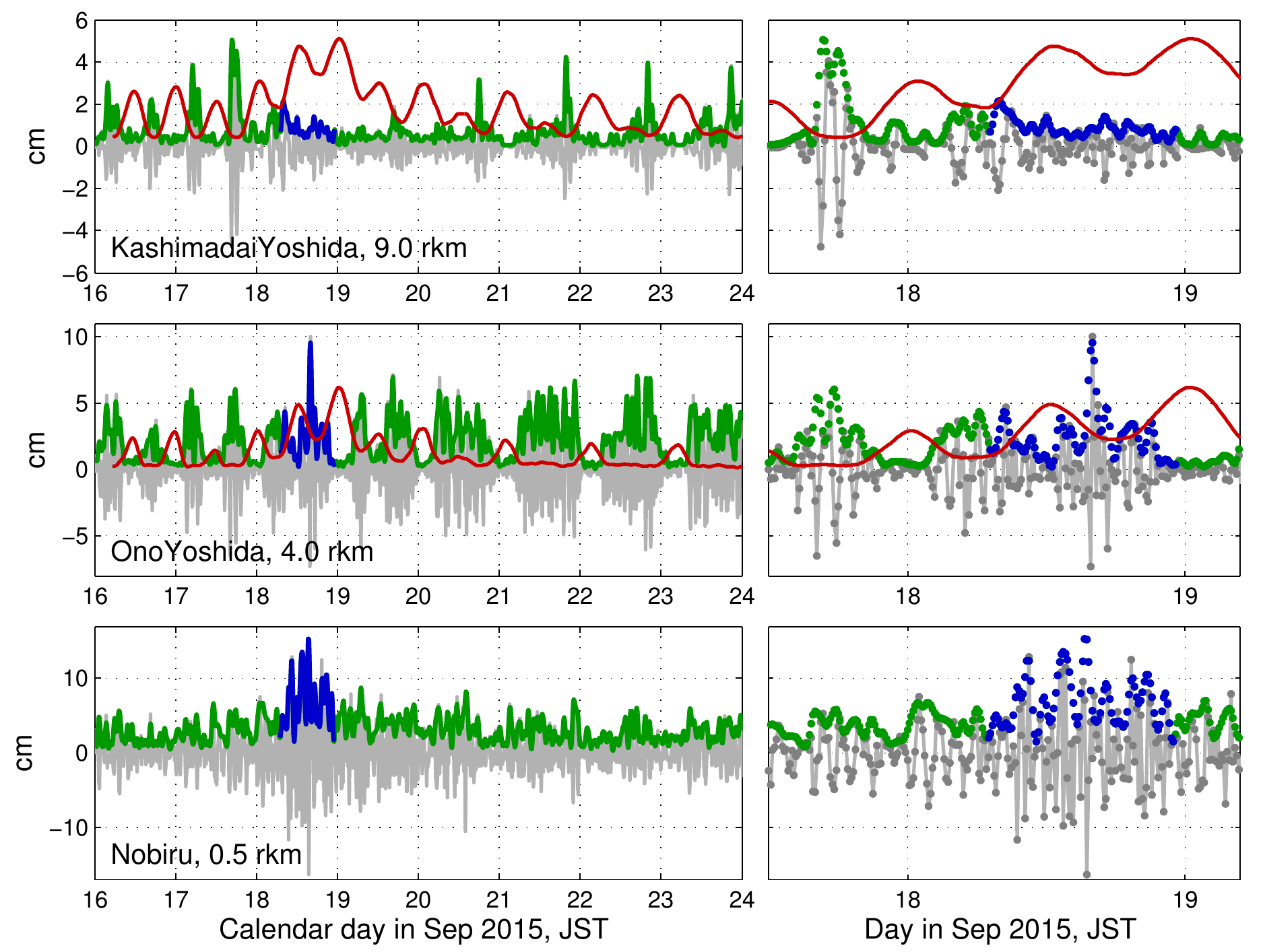}}  
	\caption{Segments of de-tided records at three stations in Yoshida River, zoomed-in on the right: de-tided measurements (gray), amplitudes of the associated wave elements (green - due to long-wave ocean noise, and blue - for 16 h after the tsunami arrival at the coast), and wave-locked slope (red curve; in cm/rkm, with a factor 0.25 for both stations) between the respective upriver station and the station near the river mouth (bottom panel).  Individual readings or wave elements are shown with dots in zoomed-in records on the right.}
	\label{wtimeY}
\end{figure}
\begin{figure}[ht]
	\resizebox{\textwidth}{!} %
		{\includegraphics{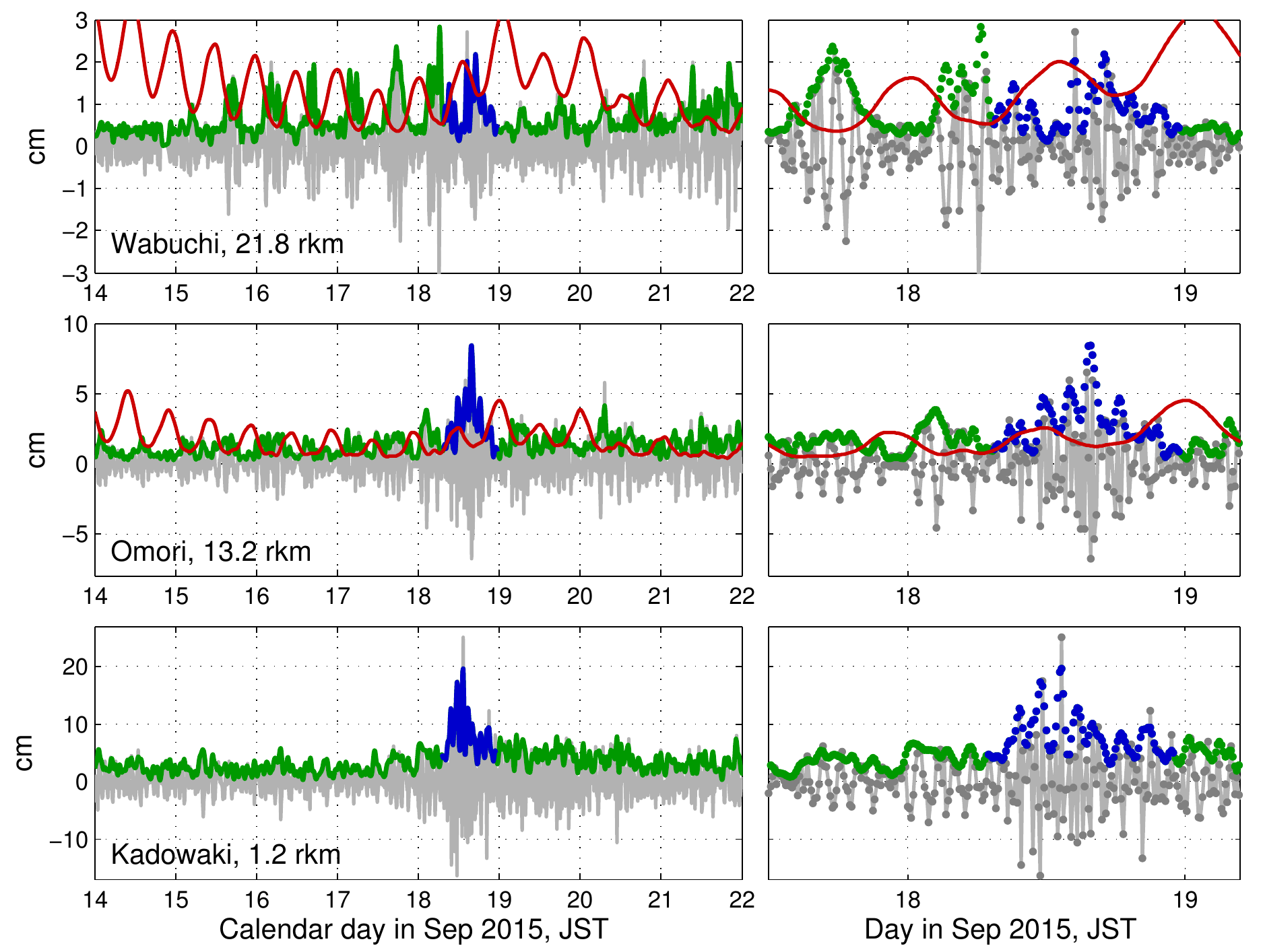}}  
	\caption{Segments of de-tided records at three stations in Old Kitakami River, zoomed-in on the right: de-tided measurements (gray), amplitudes of the associated wave elements (green - due to long-wave ocean noise, and blue - for 16 h after the tsunami arrival at the coast), and wave-locked slope (red curve; in cm/rkm, with a factor 0.5 in the upper panes, and 1.0 in the middle panes)) between the respective upriver station and the station near the river mouth (bottom panel).  Individual readings or wave elements are shown with dots in zoomed-in records on the right.}
	\label{wtimeOK}
\end{figure}
\begin{figure}[ht]
	\resizebox{\textwidth}{!} %
		{\includegraphics{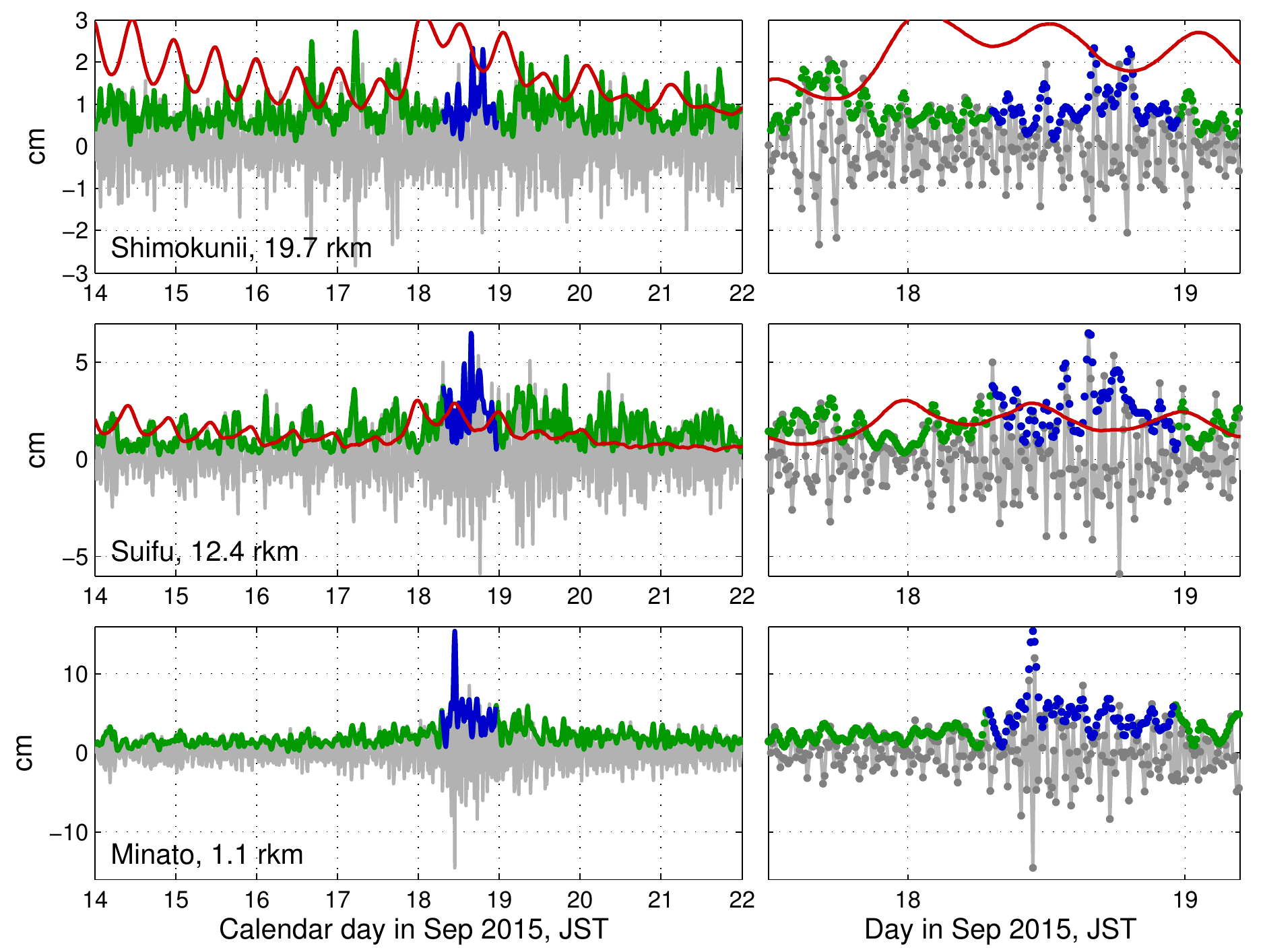}}  
	\caption{Segments of de-tided records at three stations in Naka River, zoomed-in on the right: de-tided measurements (gray), amplitudes of the associated wave elements (green - due to long-wave ocean noise, and blue - for 16 h after the tsunami arrival at the coast), and wave-locked slope (red curve; in cm/rkm, with a factor 0.5 in the upper panes, and 1.0 in the middle panes) between the respective upriver station and the station near the river mouth (bottom panel).  Individual readings or wave elements are shown with dots in zoomed-in records on the right.}
	\label{wtimeNK}
\end{figure}
\begin{figure}[ht]
	\resizebox{\textwidth}{!} %
		{\includegraphics{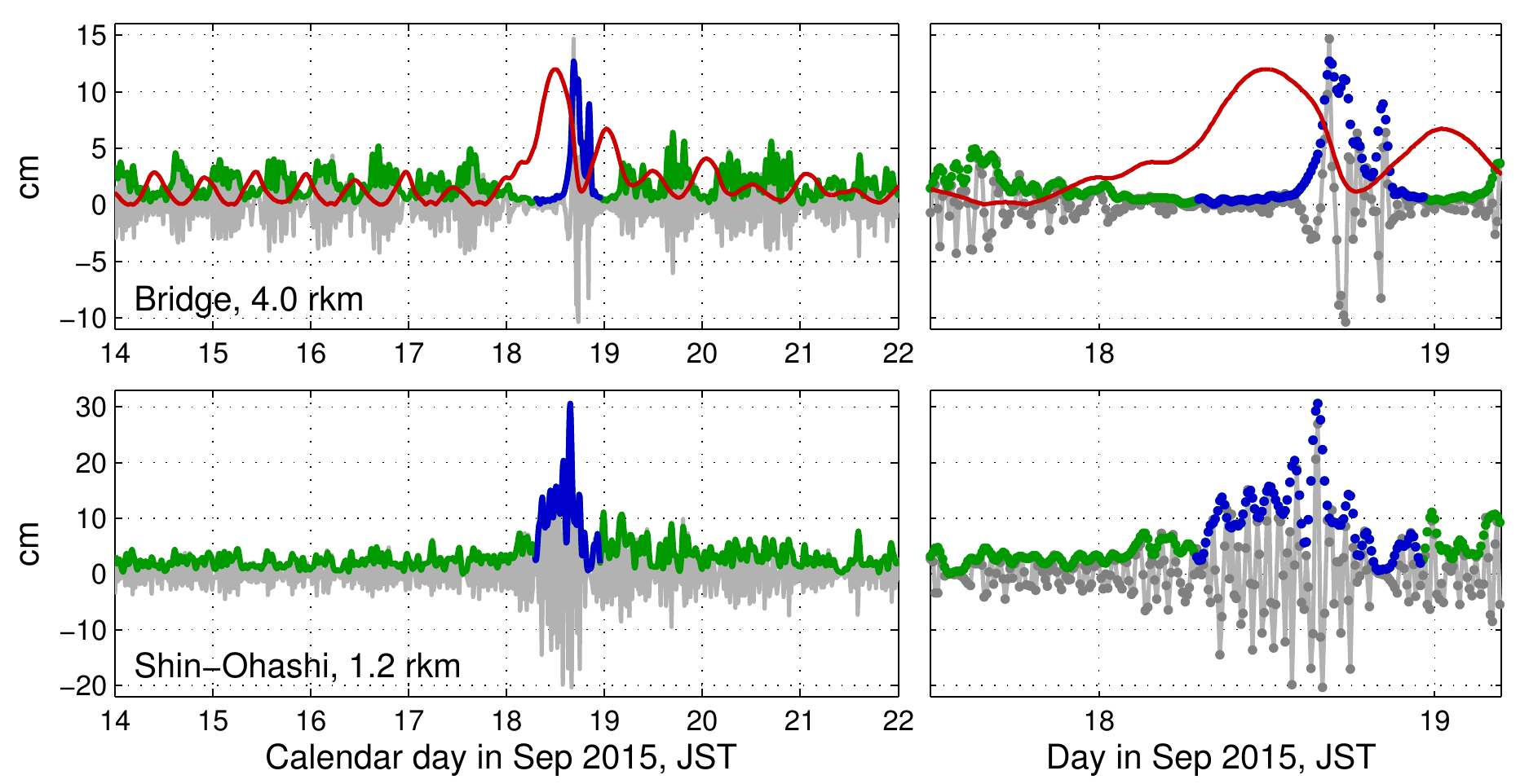}} 
	\caption{Segments of de-tided records at two stations in Mabechi River, zoomed-in on the right: de-tided measurements (gray), amplitudes of the associated wave elements (green - due to long-wave ocean noise, and blue - for 16 h after the tsunami arrival at the coast), and wave-locked slope (red curve; in cm/rkm, with a factor 0.25) between the upriver station and the station near the river mouth (bottom panel).  Individual readings or wave elements are shown with dots in zoomed-in records on the right.}
	\label{wtimeMb}
\end{figure}

Figures \ref{wtimeNz}-\ref{wtimeMb} show the records at 12 stations in 5 rivers without the longer-period (tidal and meteorological) variations, and the respective time histories of the instant wave amplitudes. De-tiding/de-trending was performed by low-pass Butterworth filtering with a 180 min cut-off period. 
The residual signal comprises the tsunami and/or coastal seiche (ocean noise).
At the downstream stations, the recorded tsunami wave is 15-30 cm high, while the ocean noise is about 5 cm high. 
Incidentally, the tsunami occurred during a high runoff in those rivers, which elevated water level at the upstream stations by about 1 m in each river, resulting in peak surface gradients, large currents, and efficient damping of the tsunami. At the upstream stations, the tsunami signal is seen only in the late afternoon of September 18th, being otherwise lower than the ocean noise on different days (see Bridge in Mabechi in Fig. \ref{wtimeMb}). 

For each upstream station, the figures \ref{wtimeNz}-\ref{wtimeMb} also include average background WLSs traveled by tsunami/noise wave elements from the most downstream station in that river. High runoff around the 18th of September brought about larger currents and larger WLS due to contribution of the riverine hydraulic gradient.  The simultaneous occurrence of the tsunami allowed to probe the larger WLS with a larger signal.
Tsunami/noise amplitude appears modulated by WLS already at the downstream stations after traveling 0.5-1.2 rkm from the mouth, with the modulation progressing upriver.
The correlation between the tsunami or the noise instant amplitude upriver and the respective background WLS is apparent in every upstream record, where an increase in the WLS traveled by the wave always coincides with the drop in the wave amplitude. Farther upriver, the intruding wave reduces to a few-hour-long trains admitted only at the lowest WLSs (see Ono in Naruse in Fig \ref{wtimeNz}; Kashimadai in Yoshida in Fig. \ref{wtimeY}; Wabuchi in Old kitakami in Fig. \ref{wtimeOK}). It was during a short decrease of the WLS in the late afternoon of September 18th, when the tsunami was able to reach the upstream stations.

It is interesting to note that WLS has two peaks per day, mostly coinciding with ebb tide, when tide is predominantly semi-diurnal; but only one peak per day for both diurnal and mixed tide, which occurs only on the transition to the lower low level. Thus no excessive damping of intruding wave (seiche or tsunami) is observed when the tide recedes to its second (higher) low level. In our records, the tide evolved from semi-diurnal before the tsunami to the mixed tide after.  Consequently, the ocean noise admitted upriver before and after the tsunami event is modulated in different ways, from two intense wave clusters (before) to a single one per day (after), as seen in top panes in Figures \ref{wtimeNz}-\ref{wtimeY}.

\section{Admittance Computations}

In an attempt to quantify the observed correlations, we compute the tsunami or ocean noise admittance factor $\kappa(\beta)$ between two stations as a function of the WLS. With the concept of wave elements traveling different surface slopes, it is expected that an element's amplitude at an upriver station $B$ (output) is proportional to its downstream amplitude at $A$ (input) with an admittance factor depending on the WLS:
\begin{equation}
\alpha_B(t)=\kappa(\beta_{AB}(t)) \cdot \alpha_A(t-\tau)
\label{kappa}
\end{equation}
In mathematical terms, space-time positions of the measurements taken at $A$ and $B$ lay on the same characteristic of the wave equation, whereas a wave element represents an elementary disturbance propagating along it. An assumption that  admittance $\kappa$ depends on ambient conditions stemmed from the background motion (tide and current combined), but does not depend on the tsunami amplitude, also implies that the propagation tract is essentially linear with respect to the tsunami component. This assumption is explain in section \ref{tsunami}.

Anticipated errors in determining a wave element amplitudes at the stations preclude calculating the admittance factor from relation \eqref{kappa} directly. These errors arise from:
\begin{enumerate}
\item
using the same propagation time $\tau$ for all wave elements; whereas (under the shallow-water approximation) wave elements propagating at different tidal phases meet different flow depth and speed and therefore travel at different celerities. 
\item
measuring $\tau$ in 10-min increments dictated by the sampling rate;
\item
aliasing in the tsunami/seiche signal from the higher frequency processes, as admitted by the instrument;
\item
de-tiding.
\end{enumerate}
At the same time, errors due to (2) are likely to offset errors due to (1). Errors due to (1) and (2) result in uncertainty in determining travel time $\tau$ for each wave element.  

De-tiding error contributes into inaccuracies in determining both tidal and tsunami components, which translates into inaccuracies in determining both WLS and instant tsunami amplitude. The 2015 tsunami had low amplitude (0.15-0.2 m in 2015, compared to 0.5 m in 2010 \cite[]{kawai} ) and relatively short period ($\thicksim 30$ min, compared to $\thicksim 1-1.5$ hr in 2010), which permits accurate extraction of the tidal component by routine filtering even at the upstream stations, where tidal spectrum widens due to nonlinearity of the propagation tract. 
Two-way (that is, applied forth and back, to ensure a zero phase shift) low-pass third-order Butterworth filter \cite[]{hamming} with a 3 h cut-off period was used to isolate tide from the shorter waves. 
With the 10-min data sampling, the filter amplitude is $H_{12}=0.9998$ at a frequency corresponding to a 12 h period; $H_{6}=0.9851$ at a 6 h period; $H_{3}=0.25$ at a 3 h period; and $H_{2}=0.0748$ at a 2 h period.  This means that a de-tiding residual left by tidal variations in the major tidal bands with, say, 50 cm amplitude would be below $50 \cdot (1-H_{12})=0.01$ cm;  the residuals left by overtides with a 5-cm amplitude would be about $5 \cdot (1-H_{6})=0.07$ cm; and the waves with a 10-cm amplitude and a 2 h period would lose no more than $10\cdot H_{2}=0.75$ cm in height after de-tiding, with much less loss for the shorter wave components. Thus 1 cm is a reasonable upper estimate for the de-tiding error.
A de-tiding error of 1 cm at one station would produce the same error in the tsunami's element amplitude, and 0.2 cm/rkm error in calculated WLS between stations separated by 5 rkm. 

\begin{figure}[ht]
\begin{tabular}{c}  
	\includegraphics[height=0.3\textwidth]{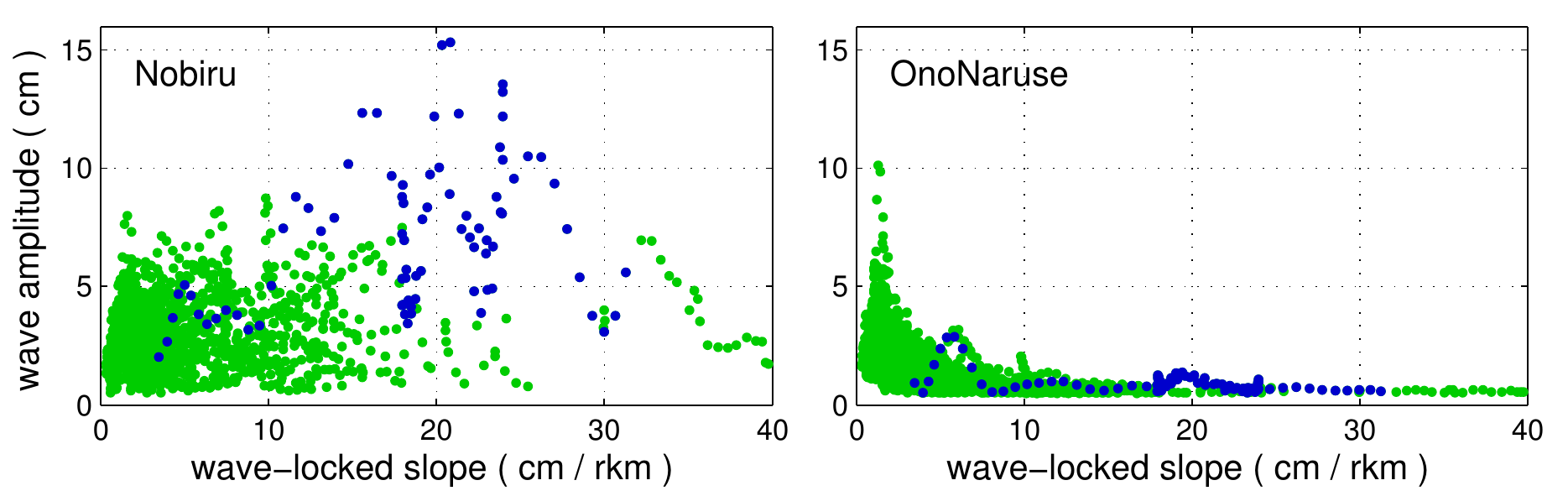} \\
	\includegraphics[height=0.7\textwidth]{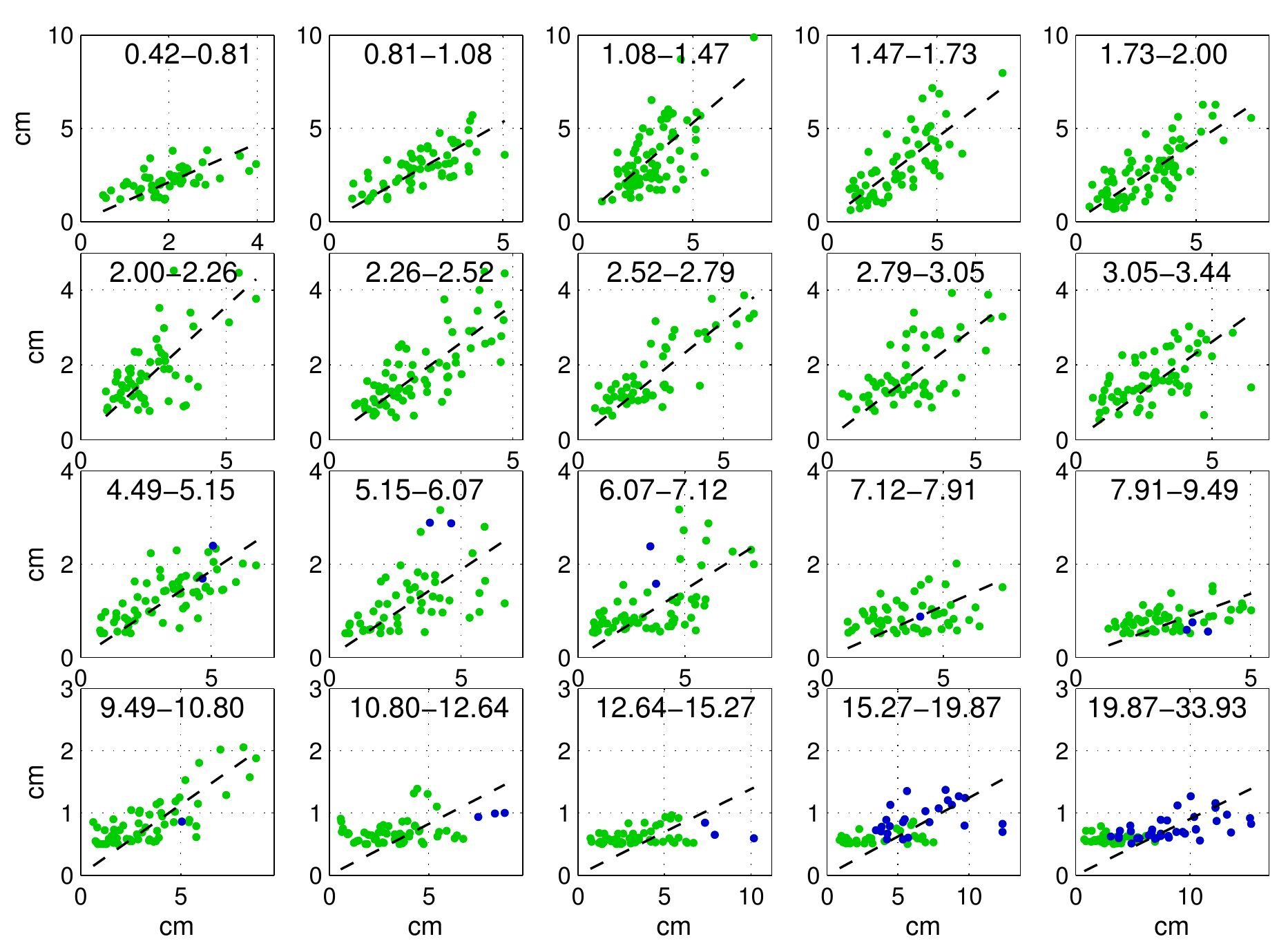}
\end{tabular}
	\caption{Top: wave elements (dots) represented by their amplitudes at Nobiru at 0.5 rkm  (left) and Ono-Naruse at 4.18 rkm (right), vs. wave-locked surface slopes traveled by each wave element between the two stations. Lower twenty panels: wave element amplitude at Ono (y-axis) vs its amplitude at Nobiru (x-axis), for all elements traveling in a range of wave-locked slopes indicated in each panel. Blue dots represent wave elements in the 2015/09 tsunami event, green dots represent long-wave ocean noise, in Naruse River.}
	\label{slpOnoN}
\end{figure}

The listed errors (1)-(4) might (or might not) result in low signal-to-noise ratios in the wave element population which will be mitigated through some simple statistical means allowing to use quantity to compensate for quality. 
The computational method is demonstrated below on an example of computing the admittance factor as a function of WLS between Nobiru and Ono stations in Naruse River. 
First, the elements with an amplitude below 0.5 cm at either station are excluded from the analysis.
Then, the wave elements are sorted according to average WLSs traveled between the stations. The elements' amplitudes at each station $\alpha_A(t-\tau)$ and $\alpha_B(t)$ are mapped against respective WLS $\beta_{AB}(t)$ in Figure \ref{slpOnoN}, top. 
We see that at Nobiru, the wave amplitudes are dictated by the ocean forcing with the highest amplitudes brought by the tsunami; whereas at Ono, the wave amplitudes are dictated by WLS with the highest amplitudes corresponding to the lower WLS. 

Next, the WLS domain was divided into uneven cells, each cell containing no less than 1/25 of the total number of wave elements in the record (2016 less the excluded elements, for a two-week no-gap record with a 10-min sampling interval). This resulted in sorting wave elements into 22 slope cells in Naruse, and into 19-23 cells in other rivers. 
Figure \ref{slpOnoN} maps wave elements within a cell by their input amplitudes (at Nobiru) and output amplitudes (at Ono). The elements in the first 10 cells and in the last 10 cells are shown, with the range of WLS for each cell displayed in the maps. 

\begin{figure}[ht]
	\resizebox{\textwidth}{!} 
		{\includegraphics{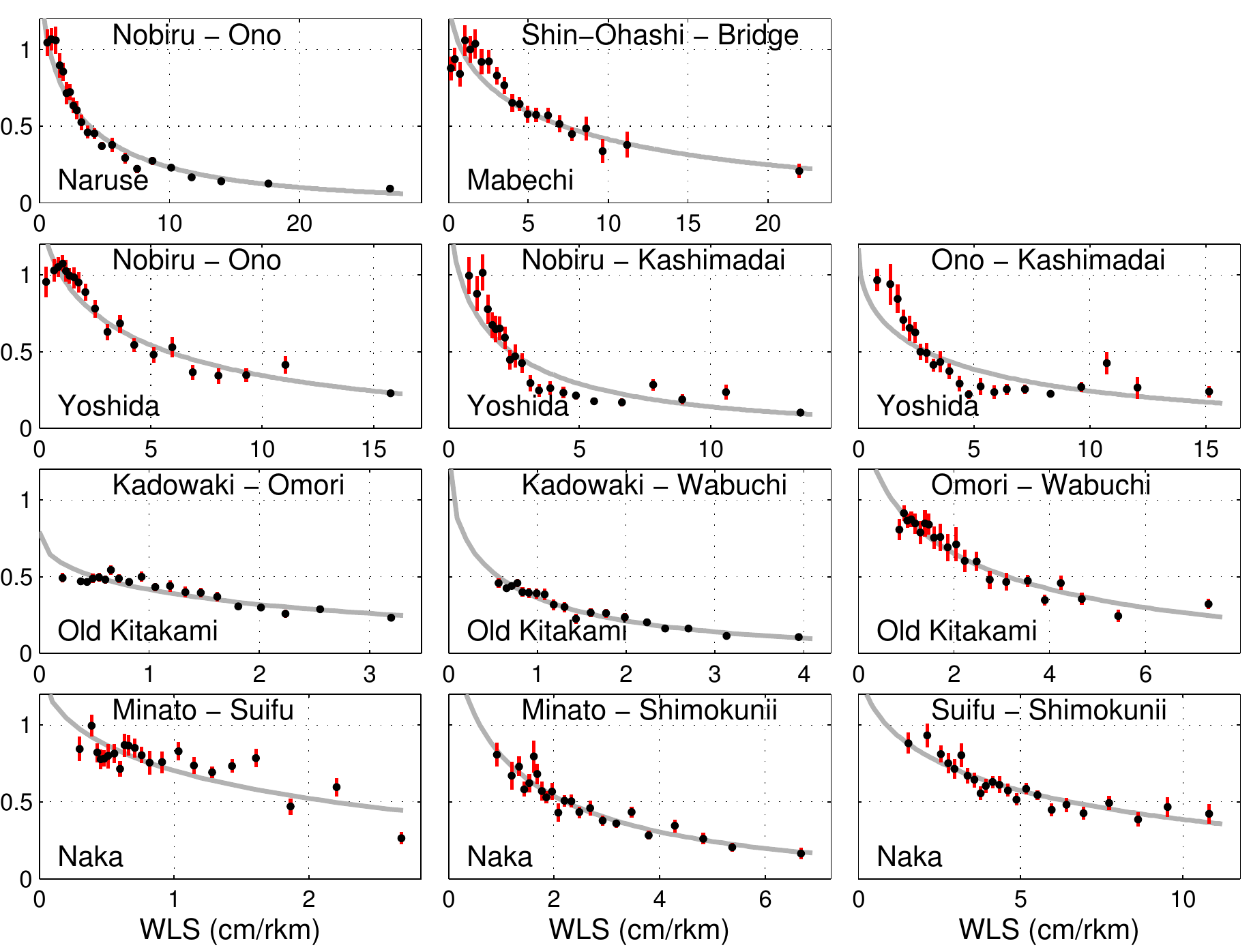}}
	\caption{Tsunami admittance factor between two stations as a function of wave-locked slope $\beta$ (black dots), fitted with a curve $a_1 \exp{(-a_2 \sqrt{\beta})}$. Red bars span 95\% confidence intervals. Respective stations and rivers are shown in the plots.}
	\label{WLS}
\end{figure}

Lastly, the dots in each cell are fitted with a straight line $\alpha_B=\kappa \cdot \alpha_A$ to find the admittance factor $\kappa$ corresponding to the mean WLS in this cell. To put an error bar on the result, the output amplitudes of $N$ wave elements in a WLS cell are considered to include white noise according to a model:
\begin{equation}
\alpha_{B,i}=\kappa \cdot \alpha_{A,i} + e_i, \ \ \ i=1, \dots, N
\label{stats}
\end{equation}
An output error $e$ consolidates all error sources discussed previously, plus an error arising from WLS uncertainty within a cell. Note, that the same or close WLSs occurred on different days, at different times, and because of freshwater discharge variations -- even at different phases of the tidal cycle. Thereby a WLS cell contains wave elements largely scattered in time, and hence the assumption of error realizations $e_i$ being statistically independent seems reasonable. Then the admittance variance can be found as
\begin{equation}
\sigma_{\kappa}^2=\sigma_e^2/{\sum{\alpha_{A,i}^2}}, 
\end{equation}
where the error variance in this WLS cell is evaluated from the data as
\begin{equation}
\sigma_e^2=\sum{(\alpha_{B,i}-\kappa \alpha_{A,i})^2}/(N-1). 
\end{equation}
The resulting function $\kappa(\beta)$ is presented in Figure \ref{WLS}, top left, with error bars spanning intervals $[\kappa-2\sigma_{\kappa}, \kappa+2\sigma_{\kappa}]$. 
This processing reveals only those wave amplitude transformations, which have occurred between Nobiru and Ono; irrespective of any input non-uniformity such as any previous transformations and/or modulation which the wave might have undergone by Nobiru. 
\clearpage

The same procedure have been repeated in every river, to obtain the wave admittance factor as a function of WLS. In Yoshida, Old Kitakami, and Naka rivers, where the tsunami was recorded at three consequent stations, the admittance factor was computed for every station pair. Typically, a river slope increases upstream, so a mean surface slope between more downstream stations $A$ and $B$ is milder than a mean slope between $A$ and the more upstream station $C$, and the latter slope is still milder than the slope between $B$ and $C$. Therefore three functions $\kappa(\beta)$ computed for these three pairs of stations represent different slope ranges and flow conditions. Wave element manipulation at other stations and rivers, analogous to that described in Naruse, is illustrated in Figures 13-22 in Supplementary Appendix II. The results of computations for all applicable station pairs, which amount to eleven sets of $\kappa(\beta)$, are displayed in  Figure \ref{WLS}. In spite of a possibly low signal-to-noise ratio in the wave element population, the developed procedure of computing admittance factors has consistently yielded fairly accurate results, with $\sigma_{\kappa} \approx 0.05 \kappa$ at least for $\kappa >0.5$. The reliability of the results has been verified by reproducing them very closely under different processing parameters, such as de-tiding with 220 min cut-off, and mapping wave elements into different number of WLS cells.

\subsection{Cross-examination of computed admittances}

\begin{figure}[ht]
	\resizebox{\textwidth}{!} %
		{\includegraphics{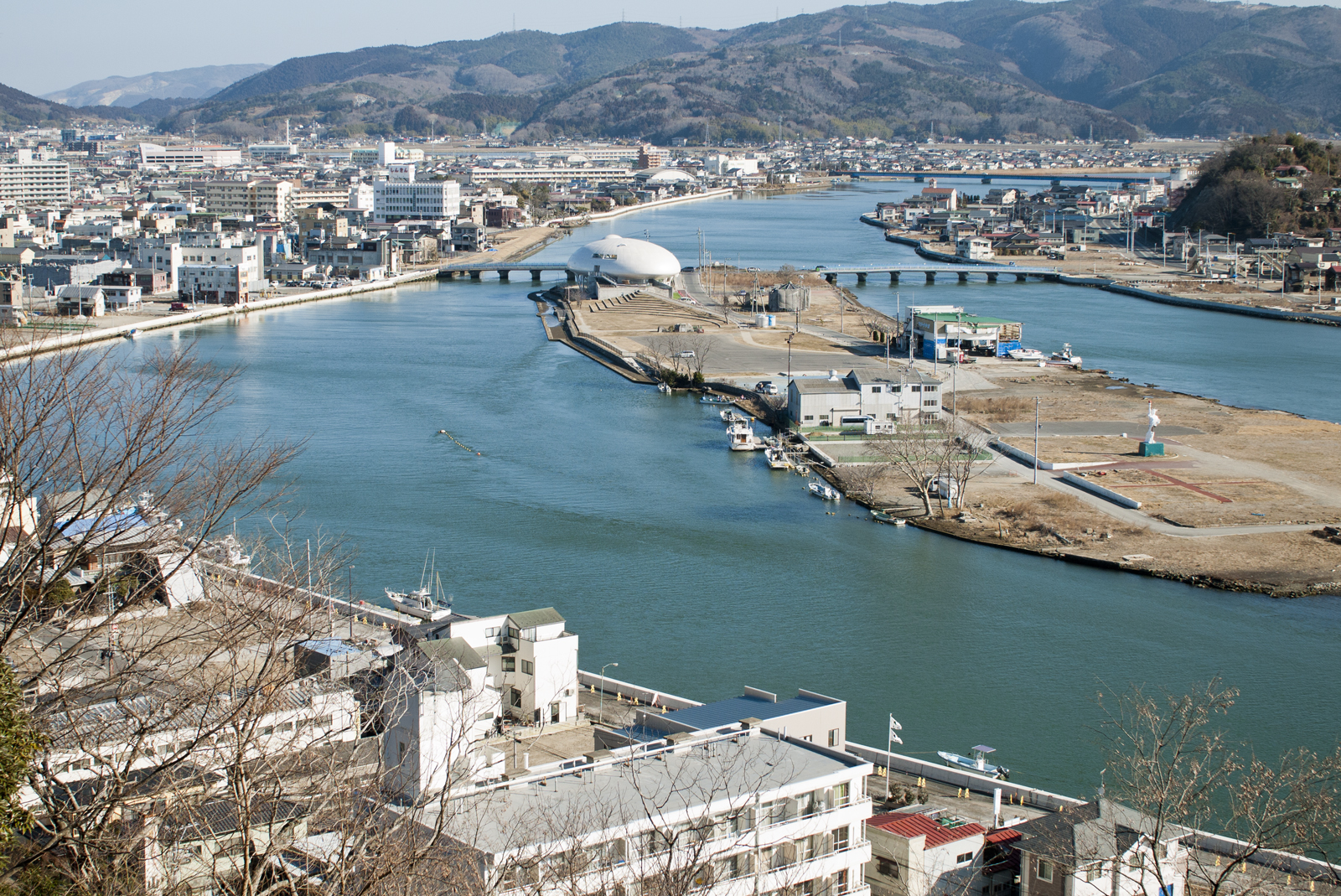}}
	\caption{Old Kitakami river between Kadonowaki and Omori gauging stations, pictured in 2013. Photo courtesy of Manabu Itoh, Japan.}
	\label{island}
\end{figure}

As seen in Figure \ref{WLS}, the admittance factor between any two stations in each river definitely correlates with the WLS, decreasing as the WLS increases (cases of seeming admittance drop at very small WLS will be discussed at the end of section \ref{physics}). 
In all discussed rivers, except Old Kitakami, the admittance at low WLS is about 100\% (admittance factor is near one). 
In Old Kitakami, the highest admittance factor between Kadonowaki and either upstream station (Omori or Wabuchi) is 0.5, whereas between Omori and Wabuchi, this factor is near one as in other rivers. Therefore, the loss of the signal is localized between Kadonowaki and Omori. The likely reason is partial reflection from an island upstream Kadonowaki, positioned in the middle of the river (see Figure \ref{island}).

\begin{figure}[ht]
\centering
	\resizebox{\textwidth}{!} 
		{\includegraphics{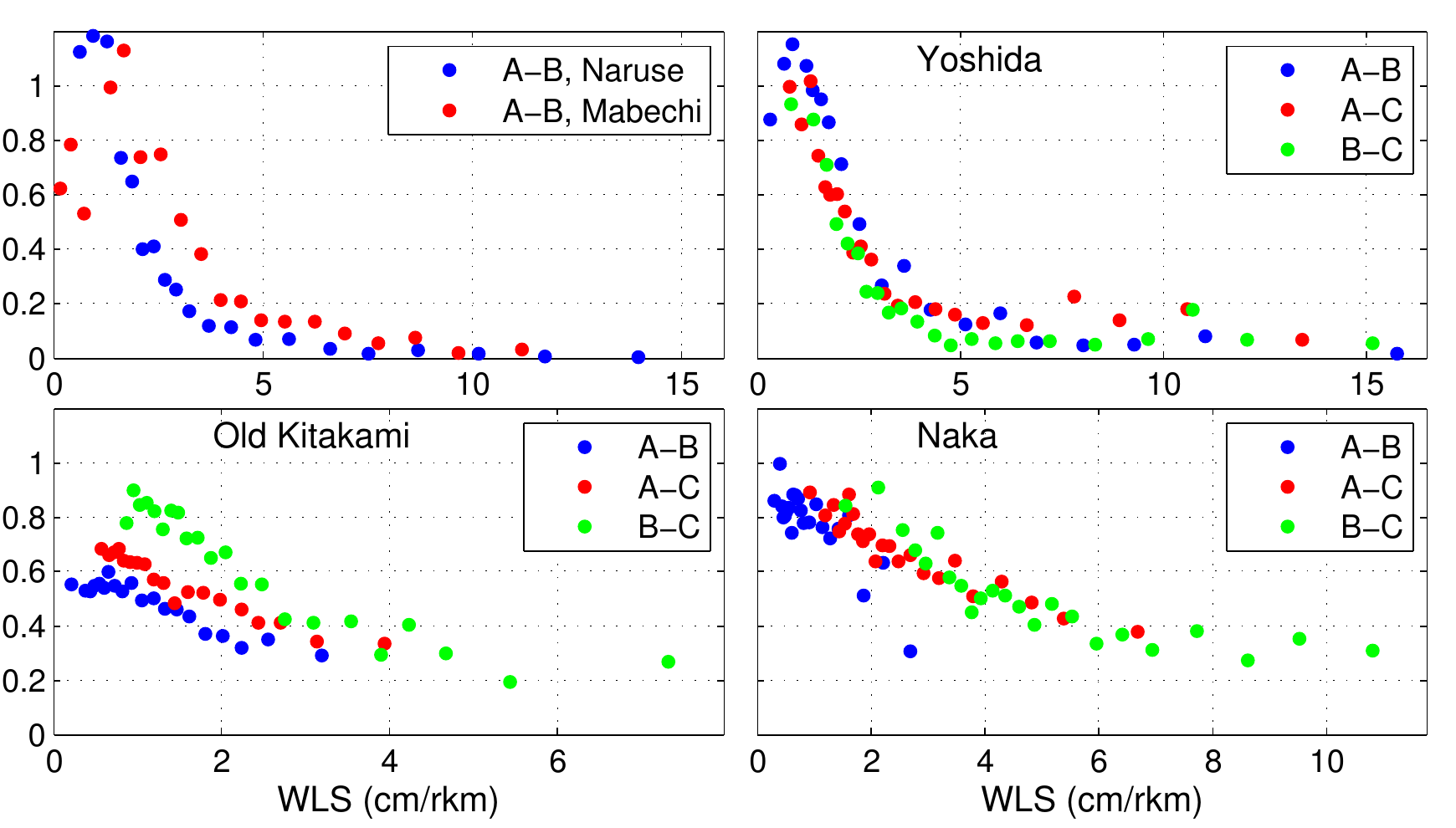}}
	\caption{Tsunami admittance factor vs WLS, recalculated to a fixed traveled distance of 10 rkm. In rivers with three stations, A refers to the most downstream gauging station, C - to the most upstream; colors refer to computations in segments AB (blue), AC (red), and BC (green).}
	\label{WLG}
\end{figure}

The distances $D$ between the gauging stations vary from 2.8 rkm to 20.5 rkm. To facilitate inter-comparisons, the admittances for each station pair, recalculated to a fixed distance of 10 rkm as $\kappa^{10/D}$, are shown in Figure \ref{WLG}, with admittances computed for different segments in the same river plotted in the same axes. 
The admittance curves appear remarkably consistent among different station pairs in the same river, and among different rivers, which offers confidence in the results. At the same WLS and over the same travel distance, admittance is higher in the larger and deeper rivers (Naka, followed by Old Kitakami). For instance, after traveling a hypothetical 10 rkm distance at 5 cm/rkm WLS, tsunami wave peak loses 50\% of its height in Naka, 75\% in Old Kitakami, 85\% in Yoshida and Mabechi, and more than 95\%  in Naruse.

As implied by the observations, a tidal river without topographic peculiarities admits small-amplitude tsunami and ocean noise with very little attenuation during small WLS. In the discussed rivers, dissipation of the shorter waves propagating atop tides is different from dissipation of tides: the latter significantly dissipate upriver, while the former might propagate virtually without losses (for as long and as far, as WLS remains low). For instance, the height of the tidal signal at Kashimadai station in Yoshida is 1/10 of its height at Nobiru (see Figure \ref{recs}), but the noise level during low WLS ($< 2$ cm/rkm, which occurs for about two-hour time once or twice per day) is as high as it was upon entering the river (Figure \ref{wtimeY}). 
Between Nobiru and Ono stations in Naruse, WLS during mixed tide remains low for as long as half a day. At the same time, the ocean noise at Ono remains as high as at Nobiru (see Figure \ref{wtimeNz}), whereas the tidal signal between these stations decreases by a factor of three. 
Between Nobiru and Ono stations in Yoshida, for most part of the day during mixed tide, WLS remains low, and the noise experiences virtually no attenuation, whereas tide loses about 1/3 of its downstream height.  

\section{Physics behind wave-locked slope}
\label{physics}
\subsection{Governing equations}
Long waves in rivers, such as tides and tsunamis, are commonly described by the cross-sectionally averaged Shallow-Water Equations (SWE) \citep{stoker}: 
\begin{eqnarray}
\label{swe1}
h_t+(hv)_x+hv \cdot b_x/b&=&0 \\
\label{swe2}
v_t+vv_x+gh_x = gd_x+f \ \ \ &OR& \ \ \ v_t+vv_x+g\eta_x=f
\end{eqnarray}
where subscript denotes partial derivatives, $x$-axis is oriented upriver with $x=0$ at the river mouth;
$v$  is the fluid velocity considered constant in each cross-section; 
$d$ is the river bed elevation counted down from the reference level;
$h$ is a cross-sectionally averaged flow depth;
$\eta$ is the free surface elevation above the reference level, so $h=d+\eta$;
$b$ is the breadth of the river considered time-invariant;
$g$ is the acceleration due to gravity; and
$f$ is the friction term expressed as 
\begin{eqnarray}
\label{kcoeff}
f=-kv|v|/h, 
\end{eqnarray}
where $k$ is the dimensionless friction coefficient.
In Manning formulation, $k = g n^2 /{h^{1/3}}$, where $n$ is the Manning roughness coefficient, typically $0.03-0.04$ $s \cdot m^{-1/3}$. For brevity, a notation $\gamma=k/h$, $f(v,h) = -\gamma v|v|$ will also be used. 

The SWE can be re-written in terms of Riemann invariants $\zeta^r$ and $\zeta^l$
\begin{equation}
\zeta^r=v+2c, \ \ \ \zeta^l=v-2c, \ \ \ c=\sqrt{gh}
\label{zetas}
\end{equation}
as a system of hyperbolic equations 
\begin{eqnarray}
\label{zetar}
\zeta^r_t+(v+c)\zeta^r_x=gd_x+f-vc \cdot b_x/b \\
\label{zetal}
\zeta^l_t+(v-c)\zeta^l_x=gd_x+f+vc \cdot b_x/b 
\end{eqnarray}
As follows from \eqref{zetas}, 
\begin{eqnarray}
\label{dzetas}
\zeta^l_t=v_t-ch_t/h, \ \ \zeta^l_x=v_x-ch_x/h, 
\end{eqnarray}   
with likewise $\zeta^r_t$ and $\zeta^r_x$. It is straightforward to check, that the equation set \eqref{zetar}-\eqref{zetal} is equivalent to the set  \eqref{swe1}-\eqref{swe2}. Subtracting \eqref{zetar} and \eqref{zetal} simplifies to the continuity equation \eqref{swe1}, while adding them simplifies to \eqref{swe2} \cite[]{diffdisp}.
A local flow state is equivalently determined by the state variables $(v,h)$ or by the Riemann invariants $(\zeta^r,\zeta^l)$. Hereafter, term `local' refers to a specific position in space and time. The Riemann invariants are said to `propagate' along corresponding characteristics: the right-going invariant $\zeta^r$ `propagates' upstream along a space-time trajectory $dx/dt=c+v$, whereas the left-going  invariant $\zeta^l$ `propagates' downstream along a trajectory $dx/dt=v-c$ (presumingly, $|v|<c$). Should, for instance, $\zeta^l$ be constant at some point $A$, then no wave motion upstream of $A$ contributes into flow state downstream of $A$. Hence a condition $\zeta^l=const$ at the river mouth implies that no fraction of the incoming tide is reflected back to the sea.

Consider a steady flow state in a river with nonuniform bottom slope, width, depth and current. By expressing  $v_x=-v(\eta_x+d_x)/h-vb_x/b$ from the continuity equation \eqref{swe1} and substituting into the momentum equation \eqref{swe2} with $v_t=0$, the steady surface slope $\eta_x$ can be expressed in terms of the local flow conditions such as the flow state $(v,h)$ and friction $f(v,h)$:
\begin{eqnarray}
\label{etax}
\eta_x=\frac{1}{1-\epsilon}\left( f/g+\epsilon d_x+(v^2/g) \cdot (b_x/b) \right), \ \ \ \epsilon=v^2/c^2
\end{eqnarray}
Apparently, for a steady flow, WLS coincides with the surface slope.
The local WLS for a right-going wave can be expressed as:
\begin{equation}
\beta= \eta_x + \eta_t/(c+v)
\label{beta}
\end{equation}
An observer locked with a specific phase in a right-going wave, while traveling a horizontal distance $\Delta x$ through time $\Delta t=\Delta x/(v+c)$, moves vertically by $\Delta y = \eta(x+\Delta x,t+\Delta x/(v+c)) - \eta(x,t) = \beta (x,t(x)) \cdot \Delta x$. Therefore, $\beta$  represents a slope of the free surface seen by this observer, consistent with WLS definition.

A few manipulations making use of relations \eqref{beta}, \eqref{dzetas}, and \eqref{zetas} follow, with $\zeta^l$ hereafter referred to as $\zeta$ (index $l$ omitted):
\begin{eqnarray}
h_t&=&(c+v)(\beta-\eta_x) \nonumber \\
hv_x&=&h(\zeta_x+ch_x/h)=h\zeta_x+ch_x \nonumber \\
h_t+(vh)_x&=&(c+v)(\beta-\eta_x)+h\zeta_x+(c+v)h_x= \nonumber \\
&=&(c+v)(\beta+d_x)+h\zeta_x
\label{man1}
\end{eqnarray}
Substituting \eqref{man1} into \eqref{swe1} and multiplying by $g/(c+v)$ simplifies to
\begin{eqnarray}
\frac{c^2}{c+v} \zeta_x+ g\beta=-gd_x-\frac{vc^2}{v+c}  \cdot \frac{b_x}{b},
\label{mm1}
\end{eqnarray}
while multiplying \eqref{zetal} by $c^2/(c^2-v^2)=1/(1-\epsilon)$ and adding to \eqref{mm1}  simplifies to
\begin{eqnarray}
\zeta_t+(1-\epsilon)g\beta=\epsilon gd_x+f+v^2 \cdot b_x/b
\label{mm2}
\end{eqnarray}
Equation \eqref{mm2} connects WLS, local flow conditions, and the variance propagating opposite to the intruding wave. 

\subsection{WLS in a purely incident wave}
If tide propagates upriver as a purely incident wave, then the outgoing Riemann invariant can experience no temporal variations, that is, $\zeta_t=0$.
Neglecting reflection is a customary practice in tidal river studies, which started with LeBlond (1978). LeBlond (1978) argued that the SWE in river environments are best approximated with a parabolic equation permitting propagation in one dimension only, which results from friction dominating the momentum balance. 
Since then, a solution for either tidal elevation or velocity in a form of a unidirectional wave decaying with upriver distance is commonly assumed and imposed \citep{godin1985, horrevoets}.
Friederichs and Aubrey (1994) applied perturbation analysis to derive the first- and second-order analytic solutions to the scaled SWE in frictional convergent tidal estuaries, and found the both solutions to represent purely incident waves, controlled by channel convergence and friction.  

Consistent with the existing views, our analytical solution for the pair $(\beta, \zeta)$ in a limiting case of a small-amplitude incident wave  in a river with constant breadth (given in Appendix) also has assumed friction as an agent responsible for the absence of the reflected wave (while the convergence factor was not included). 
More specifically, our solution procedure considered two conditions to keep $\zeta_t$ small: firstly, it is zero at the entrance; secondly, the dissipation length $L$ is much smaller than the wavelength. 
The physics behind the second condition is that with no outgoing wave at the mouth (and no discontinuities in $d$ such as a dam), formation of the reflected wave inside the river would take a channel length comparable with a wavelength, but the intruding wave dissipates over much shorter distance, and thereby the level of the reflected signal remains low. The first condition, however, is external to the river system, and was not mentioned in the previous studies. It implies that no river segment is incorporated into a larger reflecting body outside the river mouth. A possible rationale might be that a river mouth even 1 km wide is still too narrow to create an along-shore distortion on the reflected tidal wavefront, and therefore the river's inside cannot be a part of a coastal reflecting system.

In the light of the above, the sought expression for WLS is given by \eqref{mm2} after omitting $\zeta_t$, resulting in:
\begin{eqnarray}
\label{solb1}
\beta=\frac{1}{1-\epsilon}\left( f/g+\epsilon d_x+(v^2/g) \cdot (b_x/b) \right),
\end{eqnarray}
subject to a negligible level of the outgoing wave inside the river $|\zeta_t| \ll |f|$, which is provided for by three conditions: (1) no outgoing wave exists at the river mouth, (2) wavelength is much greater than the dissipation scale, and (3) not yet found limitation on river breadth variations. 

The similarity between expressions \eqref{solb1} for dynamically changing WLS  and \eqref{etax} for a steady surface slope is both appealing and explanatory. Surface slope in a steady flow acts along a trajectory of water particles traveling with the flow. Wave-locked slope acts along a space-time trajectory traveled by a sequence of fluid particles transferring the wave momentum. Thereby WLS in a unidirectional wave becomes a counterpart of the surface slope in a hypothetical  steady flow with local flow parameters of the combined tide and current. 
In particular, WLS measures dynamically-changing friction under an intruding wave, likewise a steady surface slope measures friction under a stationary flow. However, a `real' surface slope creates pressure gradient which compensates for frictional losses and maintains the flow, whereas WLS does not entirely translate into pressure, and therefore does not prevent the wave from dissipating.   

\subsection{Flow estimates with WLS}
Expressing the friction term \eqref{kcoeff} as 
$
f/g=-p_{\pm} k \epsilon$, where $p_{\pm}= sign(v)= \pm 1$, and $v^2/g=\epsilon h$,
equation \eqref{solb1} can be solved for $\epsilon=v^2/c^2$:
\begin{eqnarray}
\label{soleps}
\epsilon=\beta / \left( \beta+d_x+hb_x/b-k \cdot p_{\pm} \right)
\end{eqnarray}
For the rivers under this study, WLS varies below 20 cm/rkm; sample bed slopes $d_x$ (note that $d_x<0$ for a raising bed) are about 50 cm/rkm in Naruse, 30 cm/rkm in Yoshida, and 20 cm/rkm in Old Kitakami; this results in
$
|\beta| < 2 \cdot 10^{-4}, \ \ \ |d_x| < 5 \cdot 10^{-4}.
$
Due to gradually diminishing depth and consequently a prismatic form of a river's vertical profile, WLS is milder than the bed slope; whereas in a hypothetical river with constant depth and a steady uniform flow, $\beta=-d_x=f/g$. 
Term $h|b_x|/b$ for the study rivers do not exceed $|d_x|$, as indicated by a modest change to a river width between the mouth and  a point at $x=h/|d_x|$ where the river bottom elevates to the mean sea level (e.g., at 9 rkm in Yoshida, and at 25 rkm in Old Kitakami). The frictional coefficient evaluated with $n=0.035$ $s \cdot m^{-1/3}$, $g=9.8$ $m/s^2$, and $h=3$ m is $k=8.3 \cdot 10^{-3}$. Thus the frictional term in \eqref{soleps} is greater by about 20 times than any of its neighbors. Omitting the latter results in
 \begin{eqnarray}
\label{solb2}
\epsilon=|\beta| /k, \ \ OR \ \ \beta=f/g
\end{eqnarray}
For a sample flow estimate, with the above parameters and $\beta=2 \cdot 10^{-4}$ (large drawdown flow in Yoshida), an average Froude number of the flow between Nobiru and Ono $\sqrt{\epsilon}$ is 0.16. A conclusion to carry forward is that in the study rivers, WLS directly approximates friction.

\subsection{Tidal Wave-Locked Slope and friction upon tsunami} 
\label{tsunami}

The anticipated frictional interaction between tsunami and the tidal river environment is described by the momentum equation:
\begin{equation}
(v+V)_t + (v+V)_x (v+V) + g (\xi+\eta)_x= f(v+V,h) .
\label{qq1}
\end{equation}
where the flow is comprised of tide-and-current $(V, \ \xi)$ and a tsunami component $(v, \ \eta)$. 
The 2015 Chilean tsunami entered Honshu rivers with an amplitude about $1/4$ that of tide, and consequently with 4 times 
smaller velocity. Therefore one can assume 
$|v|<|V|$ for this small-amplitude tsunami, not to mention the ocean noise, during most of the tidal cycle. Then $sign(V+v)=sign(V)$, and the friction term can be expanded using: 
\begin{equation}
-|V+v|(V+v) =  -\frac{|V|}{V} (V+v)^2= - |V|V-2 |V| v+O(v^2)
\label{qq2}
\end{equation} 
Hereafter, we neglect the second-order terms with respect to the tsunami velocity $O(v^2)$ in all equations, as well as a contribution of tsunami-induces flow depth variations in factor $\gamma$ \eqref{kcoeff}.

Separation of tide and tsunami is done by filtering. Applying a hypothetical low-pass filter (with, say, a 3-hour cut-off period, which would also admit over-tides) to both sides of \eqref{qq1} yields an equation for the isolated tidal component:
\begin{equation}
V_t + V_x V + g \xi_x= - \gamma |V| V
\label{qq3}
\end{equation}
The residual equation describes the tsunami component:
\begin{equation}
v_t + v_x V + g \eta_x= - 2 \gamma |V|v-V_x v.
\label{qq4}
\end{equation}
As seen from equations \eqref{qq3}-\eqref{qq4}, under the condition of $|v|<|V|$, a low-amplitude tsunami has no effect on tide, whereas tide modifies tsunami celerity (equating it with its own in co-propagation) and alters conditions for tsunami dissipation. A quick estimate shows that the second term on the right side of \eqref{qq4} can be neglected compared with the first term: 
\[
\frac{V_x}{2\gamma |V|} \thicksim \frac{h^{4/3}}{2gn^2} \cdot \frac{1}{L}  \thicksim 0.024
\]
where $n=0.035$ $s \cdot m^{-1/3}$, $h=5$ m, and the dissipation length $L=15$ km were used.

In the light of the previous discussions, friction under a particular tidal phase can be estimated with a local WLS:
\begin{equation}
f(V,h)=-\gamma \cdot |V| \cdot V \approx g \beta,
\label{qq5}
\end{equation}
hence 
\begin{equation}
\gamma \cdot |V| \approx \sqrt{\gamma g |\beta|}.
\label{qq6}
\end{equation}
Then friction experienced by the tsunami becomes
\begin{equation}
- 2 \gamma |V|v \approx -2\sqrt{\gamma g |\beta|} \cdot v .
\label{eq21}
\end{equation} 
With \eqref{eq21}, we have arrived at the conclusion that
\begin{enumerate}
\item
friction experienced by a relatively small tsunami in presence of a large tidal wave and/or current exceeding tsunami current is predominantly linear with respect to the tsunami component (while friction upon the total flow is quadratic with respect to the total flow velocity). This explains the linear equation \eqref{kappa} between tsunami element amplitudes confirmed by the observations.
\item
the coefficient of the linear friction experienced by a particular tsunami wave segment is determined by the wave-locked slope under the co-propagating tidal segment. This explains the correlation between the tsunami attenuation rate and the tidal WLS observed in Figures \ref{WLS} and \ref{WLG}.  
\end{enumerate}
Expression \eqref{eq21} suggests that the admittance factor reduces exponentially with respect to $\sqrt{|\beta|}$, excluding small $\beta$. 
Observations do not contradict this suggestion, for a curve  $\kappa(\beta)=a_1 \exp{(-a_2 \sqrt{\beta})}$ provides a reasonably good fit to admittance-slope pairs obtained from the field data (Figure \ref{WLS}).

Expression \eqref{eq21} also points to a source of inaccuracy when interpreting the results obtained from the measurements. 
Water level records along a river permit to calculate an average WLS \eqref{betaAB} between stations $A$ and $B$, which relate to the local WLS as
\begin{equation}
\beta_{AB} \cdot (x_B-x_A)=\int_{x_A}^{x_B}{\beta dx},
\label{betaaver}
\end{equation}
with the integration performed along space-time wave path. Should $\beta$ change its sign as the wave travels from $A$ to $B$,
an average WLS would underestimate an absolute WLS between the stations. This happens under a tidal crest which coincides with an inflow current at $A$ and an outflow (due to the riverine flow contribution) current at $B$. Then frictional losses corresponding to the higher local WLSs will be mapped to very small average WLSs. This effect might create an artificial drop of the measurement-derived admittance factor at small WLS, which is indeed observed in some cases in Figures \ref{WLS},\ref{WLG}. 
 
\section{Conclusions} 

Our previous thesis -- that in tidal rivers, tide modulates tsunami in a certain way common to all rivers \cite[]{tolkova2015}, -- has been verified with the records containing the 2015 Chilean tsunami at 13 stations in 5 rivers on the Honshu east coast - an area in Japan most susceptible to trans-Pacific tsunamis.  
The new extensive data set, and a new methodology for data processing allowed to conclude that systematically observed correlation of the tsunami attenuation rate with tidal phase is in fact correlation with the tidally-modified wave-locked slope of the river surface.

It was deduced from the observations that 
a relatively small tsunami or ocean noise traveling at mild WLSs can propagate virtually without losses; though at the higher WLSs, combined tidal and riverine current efficiently damps the shorter waves. 
For instance, the height of the tidal signal at Kashimadai station in Yoshida is 1/10 of its height at Nobiru, but the ocean noise during low WLS ($< 2$ cm/rkm, which occurs for about two-hour time once or twice per day) is as high as it was upon entering the river.
Between most stations in the discussed rivers, tsunami/ocean noise admittance at low WLSs is about 100\%, 
whereas tidal ranges reduce upstream significantly or, as least, consistently. 
A wave in a river is dissipated through a nonlinear bottom friction, which includes frictional resistance to the wave's own motion, and its frictional interaction with other flow components.
Different dissipation conditions for tides and the shorter waves propagating atop tides arise from the fact, that the former dissipate through the frictional interaction with a given riverine current; whereas the latter interact with the combined riverine and tidal currents. Therefore different tsunami wave crests experience 'rivers' with different mean currents, including those with smaller currents and lower friction, as indicated by their milder WLSs. 

It was found analytically under the fully-nonlinear SWE framework, that WLS of a purely incident wave propagating in a river with arbitrary morphology relates to the local bottom stress in the same way as a steady surface slope does for a stationary flow, subject to neglecting reflection as described by equation \eqref{mm2}.
It was shown, that
friction experienced by a relatively small tsunami in the presence of large tidal wave and/or current is predominantly linear with respect to the tsunami flow component, with the friction coefficient being determined by the local background current which, in the study rivers, can be estimated with the tidal WLS. This explains the observed correlation between the tsunami attenuation rate and the WLS in the co-propagating tidal segment. 
In addition to defining conditions for tsunami propagation upriver, the concept of WLS might find uses in hydraulic engineering, since it might allow to deduce frictional properties of river beds or flow velocities in tidal rivers directly from water level records along the river.

\section{Acknowledgments}
The water level data in rivers in Japan and the gauging station information were obtained from the hydrology database of the Ministry of Land, Infrastructure, Transport and Tourism (MLIT) of Japan. Author sincerely thanks two anonymous reviewers for prompt responses with detailed comments and suggestions that helped to improve the manuscript.

\section{Appendix. Analytical solution in a limiting case}

After multiplying \eqref{mm1} by $1-\epsilon=(c-v)(c+v)/c^2$, and introducing notations
\begin{equation}
\tilde{\beta}=(1-\epsilon)\beta-\epsilon d_x, \ \ \ \tilde{c}=c-v,
\label{wls1}
\end{equation}
equations \eqref{mm1} and \eqref{mm2} can be re-written as 
\begin{eqnarray}
\label{newsw1}
\zeta_t+g\tilde{\beta}&=&f+v^2 \cdot b_x/b \\
\label{newsw2}
\tilde{c} \zeta_x+ g\tilde{\beta}&=&-gd_x-\tilde{c}v \cdot b_x/b
\end{eqnarray}
Here, the set \eqref{newsw1}-\eqref{newsw2} is used to express $\beta$ and $\zeta$ as functions of the local flow state for a small-amplitude wave intruding into a river with a constant breadth, so that $\tilde{c}$ can be approximated with $\sqrt{gh_0}-v_0$ and considered time-invariant, where $h_0$ and $v_0$ are the mean flow depth and speed, not necessary uniform. 
Equations \eqref{newsw1}-\eqref{newsw2} with $b_x=0$ can be solved iteratively, by applying them in an alternate manner to compute $n$-th iteration $\tilde{\beta}^{(n)}$ given $(n-1)$-th iteration $\zeta^{(n-1)}$ with  \eqref{newsw1}, and then to obtain $\zeta^{(n)}$ with $\tilde{\beta}^{(n)}$ from \eqref{newsw2}.
Starting with $\zeta^{(0)}=\zeta_0=const$, the iteration sequence is: 
\begin{eqnarray} \nonumber
g\tilde{\beta^{(1)}}&=&f \\ \nonumber
\zeta^{(1)}&=&\zeta_0- \int_0^x{(1/{\tilde{c}})(f+gd_x) dx} \\ \nonumber
g\tilde{\beta^{(2)}}&=&f+ \int_0^x{ (1/{\tilde{c}}) f_t dx} \\ \nonumber
\zeta^{(2)}&=&\zeta_0- \int_0^x{(1/{\tilde{c}})(f+gd_x) dx}-\int_0^x {(1/{\tilde{c}})dx \int_0^x{(1/{\tilde{c}})f_t dx}} \\ \nonumber
g\tilde{\beta^{(3)}}&=&f+ \int_0^x{(1/{\tilde{c}})f_t dx} + \int_0^x {(1/{\tilde{c}})dx \int_0^x{(1/{\tilde{c}})f_{tt} dx}} \\ \nonumber
\dots
\end{eqnarray}
A condition $\zeta(0,t)=\zeta_0$ implying no outgoing wave at the mouth was used in the solution process.
It can be seen, that the solution for WLS is given by an infinite series
\begin{equation}
g\tilde{\beta}=f+\Phi[f]+\Phi^2[f]+ \dots + \Phi^n[f]+ \dots
\label{sol}
\end{equation}
subject to the series convergence, where a functional operator $\Phi$ is defined as
\begin{equation}
\Phi[f]=\int_0^x{(1/{\tilde{c}}) f_t dx}
\label{phi}
\end{equation}
To evaluate the convergence, we note that since temporal variations of friction are induced by the intruding wave, concentrated within the wave-affected reach, and diminish to zero as the wave dissipates, then
\begin{equation}
|\Phi[f]| \thicksim \frac{1}{\tilde{c}}\cdot \frac{L}{T} \cdot |f| \thicksim \frac{L}{\lambda} \cdot |f|
\label{phi1}
\end{equation}
where $| \cdot |$ refers to a typical magnitude, $L$ is the wave dissipation scale, $T$ is the wave period, and $\lambda$ is the wavelength. Assuming $T=12$ h, $h=5$ m and omitting the mean current accounts for $\lambda= 300$ km (note that $cT<\tilde{c}T$ since $v_0<0$). Tide in such river is likely to reduce in amplitude at least in half after traveling, say, $L=15$ km (see Figure \ref{recs} for examples of the tidal range changes along rivers). These numbers result in $|\Phi[f]| \thicksim  |f|/20$, which implies very rapid convergence of the series \eqref{sol}. Therefore, WLS can be approximated by only the first term in the series: 
\begin{equation}
\tilde{\beta}=f/g \ \ OR \ \ \beta=\left( f/g+\epsilon d_x \right) / \left( 1-\epsilon \right)
\label{solb}
\end{equation}
Expression \eqref{solb} could had been obtained directly from \eqref{newsw1} by merely neglecting $\zeta_t$ by comparison with $f$. Indeed, for the obtained solution to be valid, $|\zeta_t| =|-\Phi[f]- \dots |\ll |f|$.


\begin{thebibliography}{99}

\bibitem[Bernier and Thompson(2007)]{bernier2007}
Bernier, N. B., and K. R. Thompson (2007), Tide-surge interaction off the east coast of Canada and northeastern United States, J. Geophys. Res., 112, C06008, doi:10.1029/2006JC003793.

\bibitem[Burwell et al.(2007)]{diffdisp}
Burwell, D., Tolkova, E., and Chawla, A. (2007), Diffusion and Dispersion Characterization of a Numerical Tsunami Model, Ocean Modelling, 19, 
10-30.

\bibitem[Friederichs and  Aubrey(1994)]{friedau}
Friederichs, C. T., and  Aubrey, D. G. (1994), Tidal propagation in strongly convergent channels, J. Geophys. Res., 99, 3321-3336.

\bibitem[Fritz et al.(2011)]{fritz2011}
Fritz, H.M., Petroff, C.M., Catal‡n, P., Cienfuegos, R., Winckler, P., Kalligeris, N., Weiss, R., Barrientos, S.E., Meneses, G., Valderas-Bermejo, C., Ebeling, C., Papadopoulos, A., Contreras, M., Almar, R., Dominguez, J.C., and Synolakis, C.E. (2011), Field Survey of the 27 February 2010 Chile Tsunami. Pure Appl. Geophys. 168(11):1989-2010, doi:10.1007/s00024-011-0283-5.

\bibitem[Godin(1985)]{godin1985}
Godin, G. (1985), Modification of river tides by discharge, J. Ports Water. Coast. Ocean Eng., 111, 257-274.

\bibitem[Godin(1999)]{godin1999}
Godin, G. (1999), The propagation of tides up rivers with special considerations on the upper Saint Lawrence river, Estuarine Coastal Shelf Sci., 48, 307Ð324.

\bibitem[Hamming(1998)]{hamming}
Hamming, R.W. (1998), Digital Filters, 3d edition (Dover Pub., Inc., Mineola, New York, NY, USA).

\bibitem[Horrevoets et al.(2004)]{horrevoets}
Horrevoets, A. C., Savenije, H. H. G., Schuurman, J. N., and Graas, S. (2004), The influence of river discharge on tidal damping in alluvial estuaries, J. Hydrol., 294, 213Ð228.

\bibitem[Kalmbacher and Hill(2015)]{hill}
Kalmbacher, K.D. and D.F. Hill (2015), Effects of Tides and Currents on Tsunami Propagation in Large Rivers: Columbia River, United States, J. Waterway, Port, Coastal, and Ocean Engineering, DOI: 10.1061/(ASCE)WW.1943-5460.0000290

\bibitem[Kayane et al.(2011)]{kayane}
Kayane, K., Min, R., Tanaka, H., and Tinh, N.X. (2011),
Influence of River Mouth Topography and Tidal Variation on Tsunami Propagation into Rivers.
Journal of JSCE, Ser.B2 (Coastal Engineering), Vol.B2-67(1), 2011, pp.I\_246-I\_250 (in Japanese)

\bibitem[Kawai et al.(2014)]{kawai}
Kawai, H., Satoh, M., Kawaguchi, K., and Seki, K. (2014), 2010 Chile and 2011 Tohoku Tsunami Profiles Measured by GPS Buoys and Coastal Wave and Tide Gauges in a Nationwide Ocean Wave Information Network for Ports and Harbors. J. Waterway, Port, Coastal, Ocean Eng., 10.1061/(ASCE)WW.1943-5460.0000235, 135-145.

\bibitem[Kowalik and Proshutinsky(2010)]{kowpro}
Kowalik, Z. and Proshutinsky, A. (2010), Tsunami-tide interactions: A Cook Inlet Case Study, Continental Shelf Research, 30, 633-642.

\bibitem[Kowalik et al.(2006)]{kowpros}
Kowalik, Z., Proshutinsky, T., and Proshutinsky, A. (2006), Tide-tsunami interactions, Sci. Tsunami Hazards, 24, 242-256.

\bibitem[LeBlond(1978)]{leblond}
LeBlond, P.H. (1978), On tidal propagation in shallow rivers, J. Geophys. Res., 83(C9), 4717-4721.

 \bibitem[Liu et al.(2013)]{liu2013}
 Liu H., Shimozono T., Takagawa T., Okayasu A., Fritz H. M., Sato S., and Tajima Y. (2013), The 11 March 2011 Tohoku Tsunami Survey in Rikuzentakata and Comparison with Historical Events. Pure Appl. Geophys. 170, 1033-1046.

\bibitem[Mori et al.(2011)]{morisurvey} 
Mori N., Takahashi T., Yasuda T., and Yanagisawa H. (2011), Survey of 2011 Tohoku earthquake tsunami inundation and runup, GRL, v.38, L00G14, doi:10.1029/2011GL049210

\bibitem[Stoker(1957)]{stoker}
Stoker, J.J. (1957), Water Waves (Interscience Pub., Inc., New York, NY, USA).

\bibitem[Tanaka et al.(2013)]{tanaka2013}
Tanaka, H., Kayane, K., Adityawan, M.B., Farid, M. (2013), The effect of bed slope to the tsunami intrusion into rivers. Proceedings of 7th International Conference on Coastal Dynamics, 1601-1610.

\bibitem[Tanaka et al.(2014)]{tanaka2014}
Tanaka, H., Kayane, K., Adityawan, M.B., Roh, M., Farid, M. (2014), Study on the relation of river morphology and tsunami propagation in rivers. Ocean Dynamics, 64(9), 1319-1332. DOI:10.1007/s10236-014-0749-y

\bibitem[Tolkova(2013)]{tolkova}
Tolkova, E. (2013), Tide-Tsunami Interaction in Columbia River, as Implied by Historical Data and Numerical Simulations. Pure and Applied Geophysics, 170(6), 1115-1126. doi: 10.1007/s00024-012-0518-0

\bibitem[Tolkova et al.(2015)]{tolkova2015}
Tolkova, E., Tanaka, H., Roh, M. (2015), Tsunami observations in rivers from a perspective of tsunami interaction with tide and riverine flow. Pure and Applied Geophysics, Vol. 172, Issue 3-4, pp. 953-968.
DOI: 10.1007/s00024-014-1017-2

\bibitem[Wilson and Torum(1972)]{wilson2}
Wilson, B.W. and Torum, A., Effects of the Tsunamis: An Engineering Study, In The Great Alaska Earthquake of 1964: Oceanography and Coastal Engineering (Committee on the Alaska Earthquake, National Research Council) (National Academy of Sciences, Washington, D.C., USA, 1972) pp. 361-526.

\bibitem[Yeh et al.(2012)]{yeh}
Yeh, H., Tolkova, E.,  Jay, D., Talke, S.,  Fritz, H. (2012),  Tsunami Hydrodynamics in the Columbia River. Journal of Disaster Research, Vol.7, No.5, 604-608.

\bibitem[Zhang W. et al.(2010)]{zhang2010}
Zhang, W.-Z., Shi, F., Hong H.-S., Shang, S.-P.,  Kirby, J. T. (2010), Tide-surge Interaction Intensified by the Taiwan Strait, J. Geophys. Res., 115, C06012, doi:10.1029/2009JC005762.

\bibitem[Zhang Y. et al.(2011)]{zhang}
Zhang, Y.J., Witter, R.C., Priest, G.R. (2011), Tide-tsunami interaction in 1964 Prince William Sound tsunami. Ocean Modelling, 40, 246-259.

\end{thebibliography}
\end{document}